\documentclass{aa}  
\usepackage[varg]{txfonts}
\usepackage{graphicx}
\usepackage{amssymb,amsmath}
\usepackage{booktabs,multirow}
\usepackage{bm}
\usepackage{natbib}
\bibpunct{(}{)}{;}{a}{}{,}

\begin{document}

\title{Energy and helicity evolution in a flux emergence simulation}

\author{K. Moraitis$^1$ \and V. Archontis$^1$ \and G. Chouliaras$^2$}
\institute{Physics Department, University of Ioannina, Ioannina GR-45110, Greece \and School of Mathematics and Statistics, St Andrews University, St Andrews, KY16 9SS, UK}

\date{Received ... / Accepted ...}

\abstract 
{}
{The main aim of this work is to study the evolution of the recently introduced relative helicity of the magnetic polarity inversion line (PIL) in a magnetohydrodynamics simulation.}
{The simulation used is a typical flux emergence simulation in which there is additionally an oblique, pre-existing magnetic field. The interaction of the emerging and ambient fields produces intense coronal activity, with four jets standing out. The 3D magnetic field allows us to compute various energies and helicities, and to study their evolution during the simulation, especially around the identified jets. We examine the evolution of all quantities in three different regions: in the whole volume, in three separate subvolumes of the whole volume, and in a 2D region around the PIL on the photosphere.}
{We find that the helicities are in general more responsive to the jets, followed by the free energy. The eruptivity index, the ratio of the current-carrying helicity to the relative helicity, does not show the typical behaviour it has in other cases, as its variations do not follow the production of the jets. By considering the subvolumes we find that the magnetic field gets more potential and less helical with height. The PIL relative helicity confirms the recent results it showed in observed active regions, exhibiting stronger variations during the jets compared to the standard relative helicity. Moreover, the current-carrying helicity around the PIL has a similar behaviour to the PIL relative helicity, and so this quantity could be equally useful in solar eruptivity studies.}
{}


\titlerunning{Energy and helicity evolution in a flux emergence simulation}
\authorrunning{Moraitis et al.}

\maketitle

\section{Introduction}
\label{sect:introduction}

Solar coronal jets are transient, energetic phenomena of the Sun that power the solar wind and are also related to coronal heating. Coronal jets are collimated plasma ejections with a characteristic inverse Y shape that are observed in extreme ultraviolet and X-rays, mostly in coronal holes. They typically occur with a frequency of 2-3 per hour, lifetimes of 10~mins, velocities around 200~km s$^{-1}$, and heights of up to a few 100~Mm \citep{jetreview}. Jets are often helical, and also show evidence of untwisting \citep{patsourakos08}. The standard picture we have for the formation of jets is that a small parasitic polarity grows inside a larger-scale, pre-existing magnetic field of the opposite polarity, and this leads to magnetic reconnection between the two systems \citep{shibata92}. Jets can be divided into standard jets and blowout jets \citep{moore10}, depending on whether the smaller bipole is eruptive or not.

The formation of coronal jets can be modelled by different types of magnetohydrodynamics (MHD) simulations. A first class of jet experiments is when magnetic flux emerges from below the photosphere and collides with a pre-existing coronal field \citep[e.g.][]{moreno13,archontis13,torok24}. Another class of experiments is when an initial magnetic configuration containing a null point is slowly driven to instability by continuous photospheric motions \citep{pariat09,karpen17,pariat23}. In both cases, the jet is produced by magnetic reconnection when the opposite-polarity fields meet.

A physical quantity that is often examined in MHD simulations in general, and in jet simulations in particular, as it is related to the twist, is magnetic helicity. The significance of magnetic helicity is evident, as it is one of the three conserved quantities of ideal MHD \citep{woltjer58}. In astrophysical applications, the appropriate form it has is given by relative helicity, which is expressed with the help of a reference magnetic field \citep{BergerF84,fa85}. Moreover, relative helicity can be uniquely split into two gauge-independent components; namely, the current-carrying helicity and the volume-threading helicity \citep{berger88}.

The ratio of the current-carrying helicity to the total relative helicity has been shown to have strong relations with solar eruptivity \citep{pariat17}, and is thus often referred to as the (helicity) eruptivity index. This quantity has been applied successfully in many simulated active regions (ARs) \citep{linan18,pariat23}, and in observed ARs as well \citep{moraitis19b,thalmann19,gupta21,green22}. The typical behaviour of the eruptivity index is that it increases until a case-dependent threshold is reached right before eruptive events, and then drops quickly.

Another helicity-related quantity that is ideal to study in MHD simulations is relative field line helicity (RFLH). This quantity can be considered as the density of relative helicity \citep{yeates18,moraitis19}, as it can highlight the locations where helicity is most important. It has been studied recently around the time of a strong flare from a single AR; namely, the X2.2 flare of AR 11158 \citep{moraitis21}. It was shown there that the morphology of RFLH indicates a flare-related decrease in relative helicity that is coincident with the flare ribbons. Additionally, the relative helicity contained in the ribbons showed the same sharp absolute decrease as the volume helicity during the flare.

In a continuation of this work with a wider AR sample, \citet{moraitis24} has shown that the relative helicity contained in a narrow region around the polarity inversion line (PIL) of the magnetic field is linked to solar flaring activity. This conclusion was based on the sharpest decreases experienced by the PIL helicity during solar flares compared to those by the other helicities that were examined. A first motivation for the current work is therefore to see whether these results hold in a simulated AR as well. It is also interesting to examine the behaviour of the current-carrying component of RFLH, something not done before, and the controlled environment of an MHD simulation is ideal for this purpose.

The main aim of this work is to examine whether the recent results for the eruptivity-indicating PIL helicity are relevant for a flux emergence MHD simulation of solar jet production. In addition to that, we also examine the overall evolution of various energy- and helicity-related quantities during the simulation. In Sect.~\ref{sect:data}, we describe the main characteristics of the MHD simulation that we use. In Sect.~\ref{sect:method}, we define all quantities of interest and the methodology for computing them from the simulation. In Sect.~\ref{sect:results}, we present the obtained results, and finally in Sect.~\ref{sect:discussion} we summarise and discuss the results of the paper.

\section{The magnetohydrodynamics simulation}
\label{sect:data}

The MHD simulation that we use is a typical flux-emergence jet experiment in which a highly twisted flux tube emerges into a stratified atmosphere that resembles the solar atmosphere \citep{archontis14,syntelis17}. The flux tube is along the $y$ axis, is inserted at a depth of $z=-2.3$~Mm, and has a magnetic field strength of 7900~G at its axis. Another ingredient of the simulation is that the partial ionisation of the plasma is taken into account, although in a single-fluid description \citep[e.g.][]{leake13}. There is additionally an oblique ambient magnetic field of strength $10$~G that makes an angle of 11$^\mathrm{o}$ with respect to the vertical direction (along the $z$ axis). Apart from the ambient field, the set-up is identical to the PI case of \citet{chouliaras23}, where more information can be found about the simulation.

\begin{figure}[ht]
\centering
\includegraphics[width=0.46\textwidth]{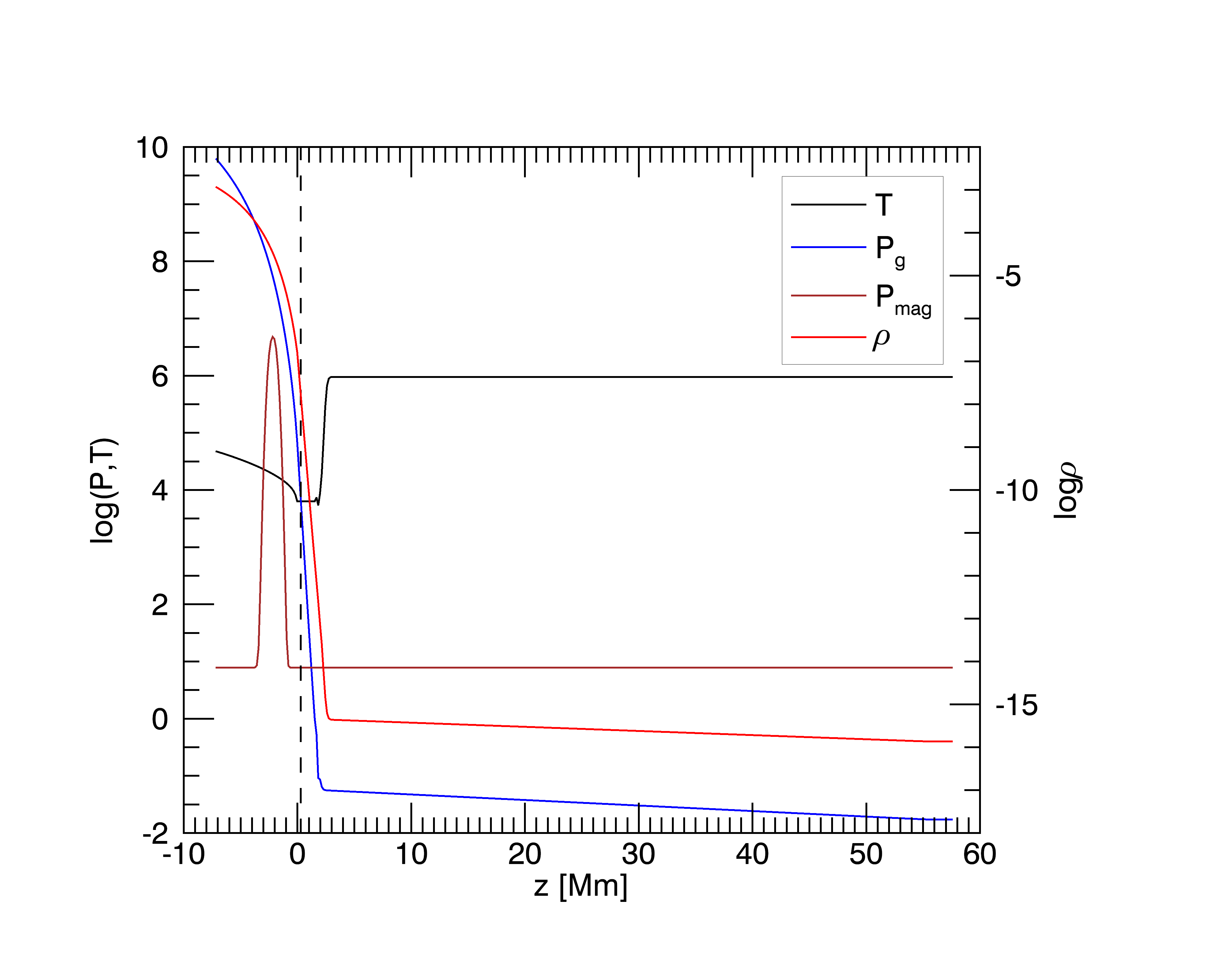}

\includegraphics[width=0.4\textwidth]{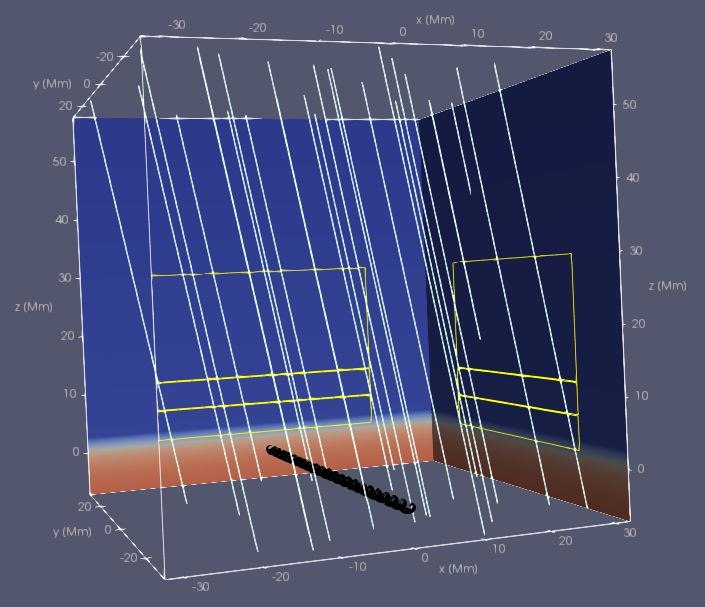}
\caption{Initial set-up of the flux emergence MHD simulation.\protect\\
(Top) Dependence on height of the initial values of density, temperature, gas, and magnetic pressure (in cgs units). The vertical line denotes the location of the photosphere.\protect\\
(Bottom) Initial configuration of the magnetic field, consisting of the twisted flux tube (black field lines) and the uniform, oblique ambient field (white field lines). The density stratification is shown with colour in the background, on a logarithmic scale. The yellow boxes are the projections of the subvolumes examined in Sect.~\ref{sect:results2} on the respective planes.}
\label{mhdsetup}
\end{figure}

The initial set-up of the simulation is depicted in Fig.~\ref{mhdsetup}. The computational box is uniform with 420 pixels in each direction that correspond to 64.8~Mm in physical dimensions. The convection zone occupies the lower 7.2~Mm below the photosphere and so the volume of interest is 64.8~Mm$\times$64.8~Mm$\times$57.6~Mm. The simulation consists of 100 snapshots in total, of which we have used 73: snapshots 15-75 with a cadence of 1, and snapshots 77-99 with a cadence of 2. Snapshots before 15 have not been examined, since the flux tube is still below the photosphere. The time unit for each snapshot is equal to 86.9~s, and so the total duration of the simulation corresponds to $\sim$145~min of real time. The full details of the simulation are discussed in an upcoming work (Chouliaras et al. 2024, in preparation), in which the comparison with the fully ionised case is also made.

The solenoidality of the produced MHD fields is of course excellent, as was expected from the divergence-preserving Lare3D code that was used to produce the simulation \citep{arber01}. As an example, the divergence energy ratio \citep[e.g.][]{valori16} takes average values of $0.01$, well below the threshold of $\sim 0.05$ that is needed for a reliable helicity estimation \citep{thalmann19a}.

\begin{figure*}[ht]
\centering
\includegraphics[width=0.24\textwidth]{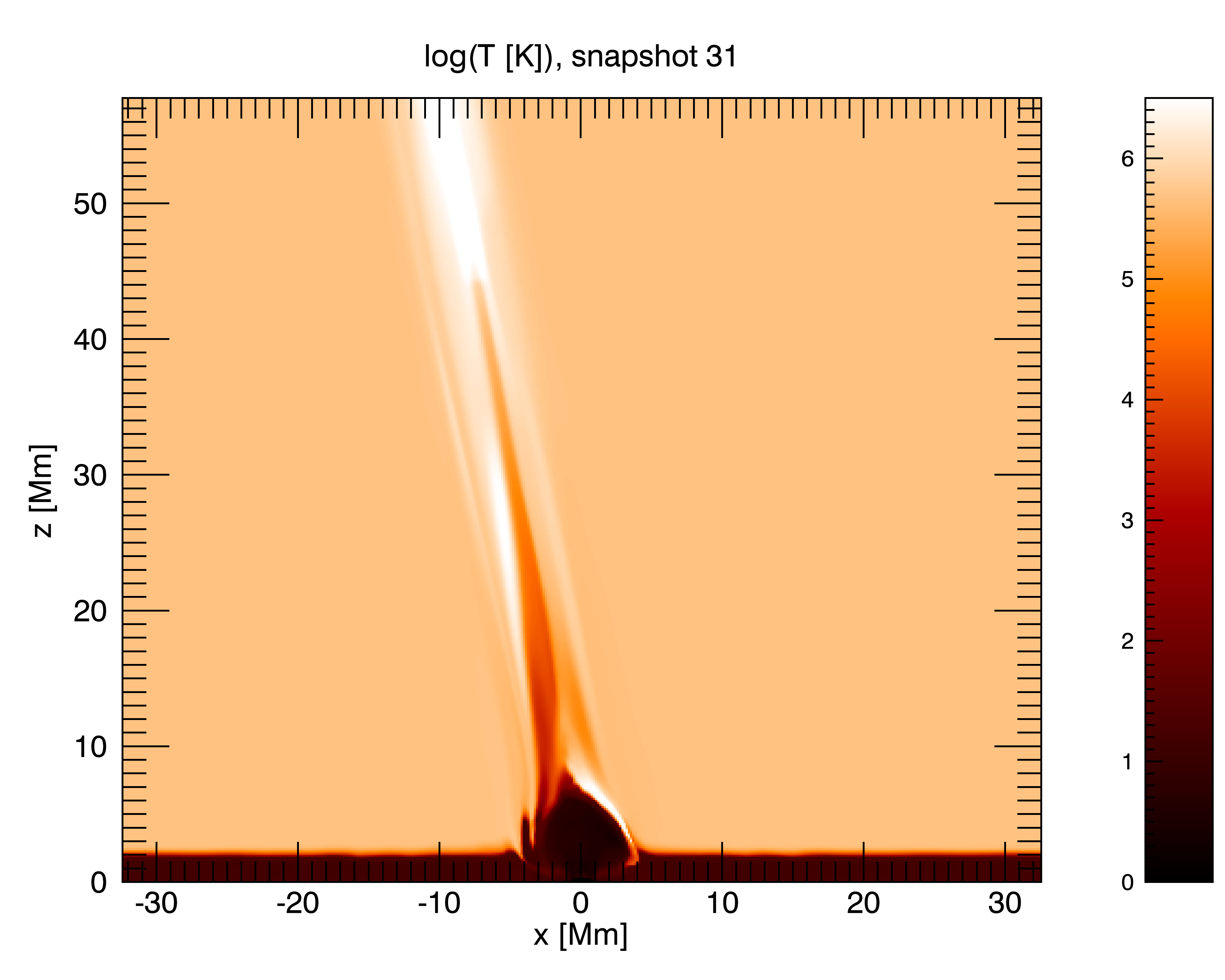}%
\includegraphics[width=0.24\textwidth]{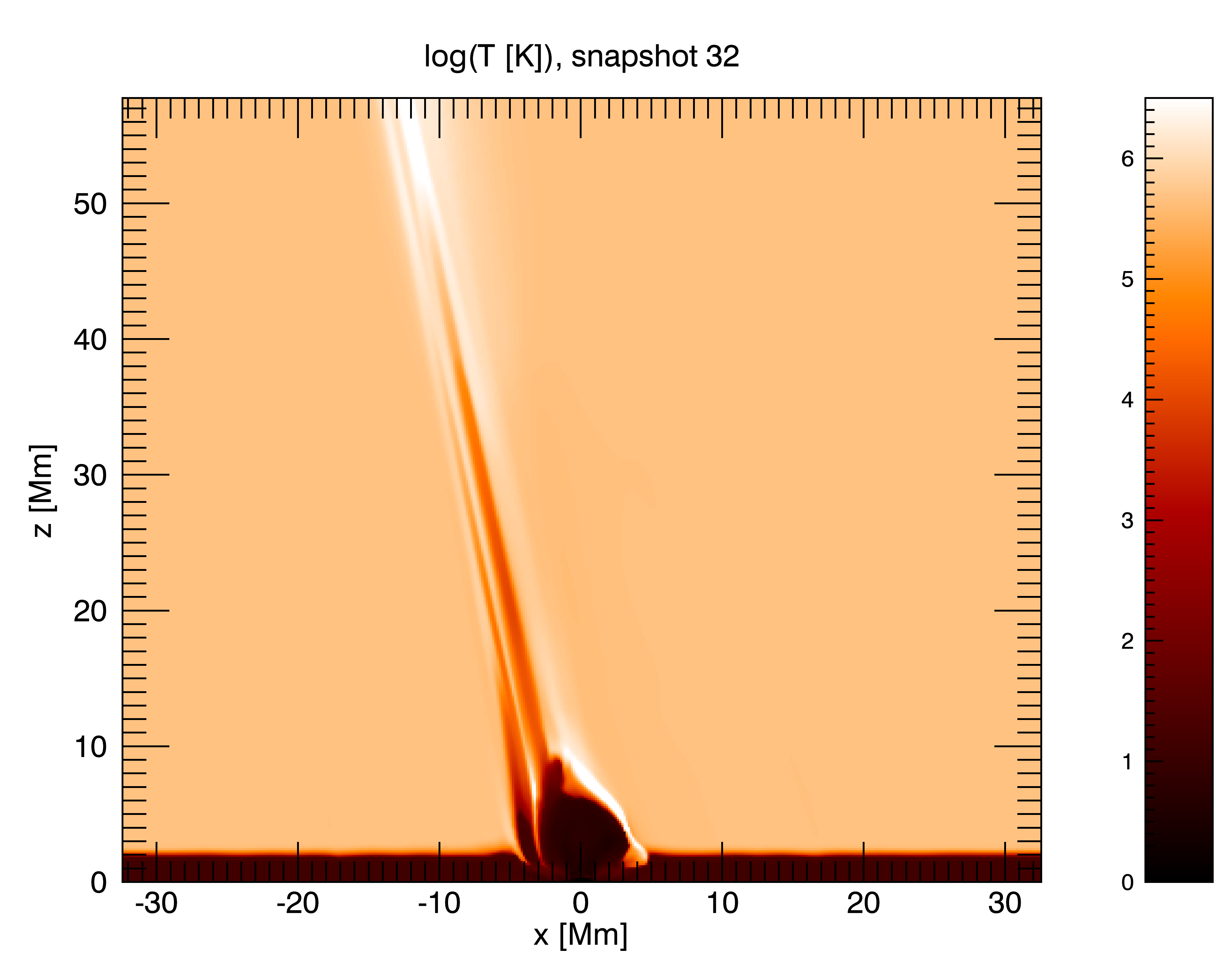}%
\includegraphics[width=0.24\textwidth]{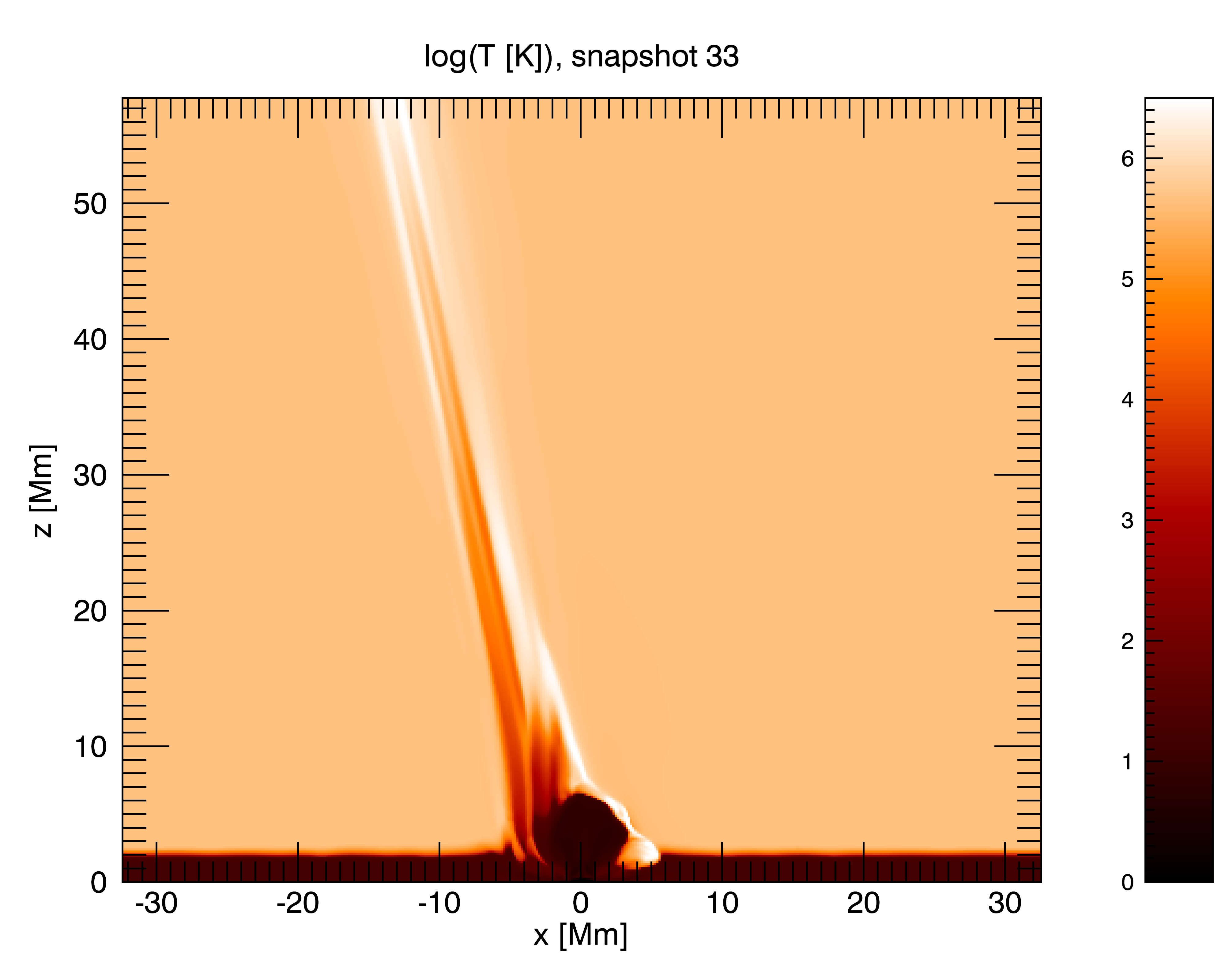}%
\includegraphics[width=0.24\textwidth]{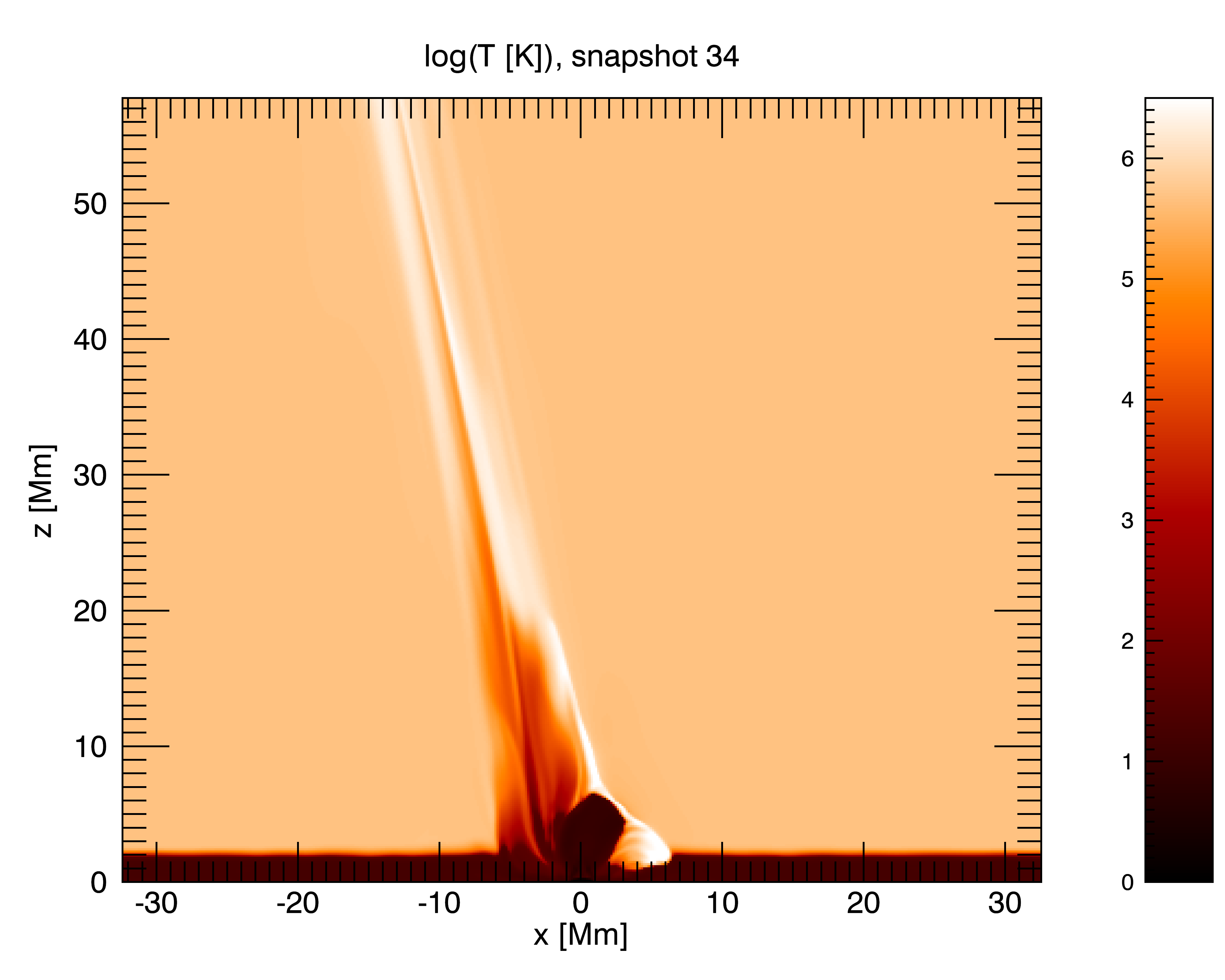}

\includegraphics[width=0.24\textwidth]{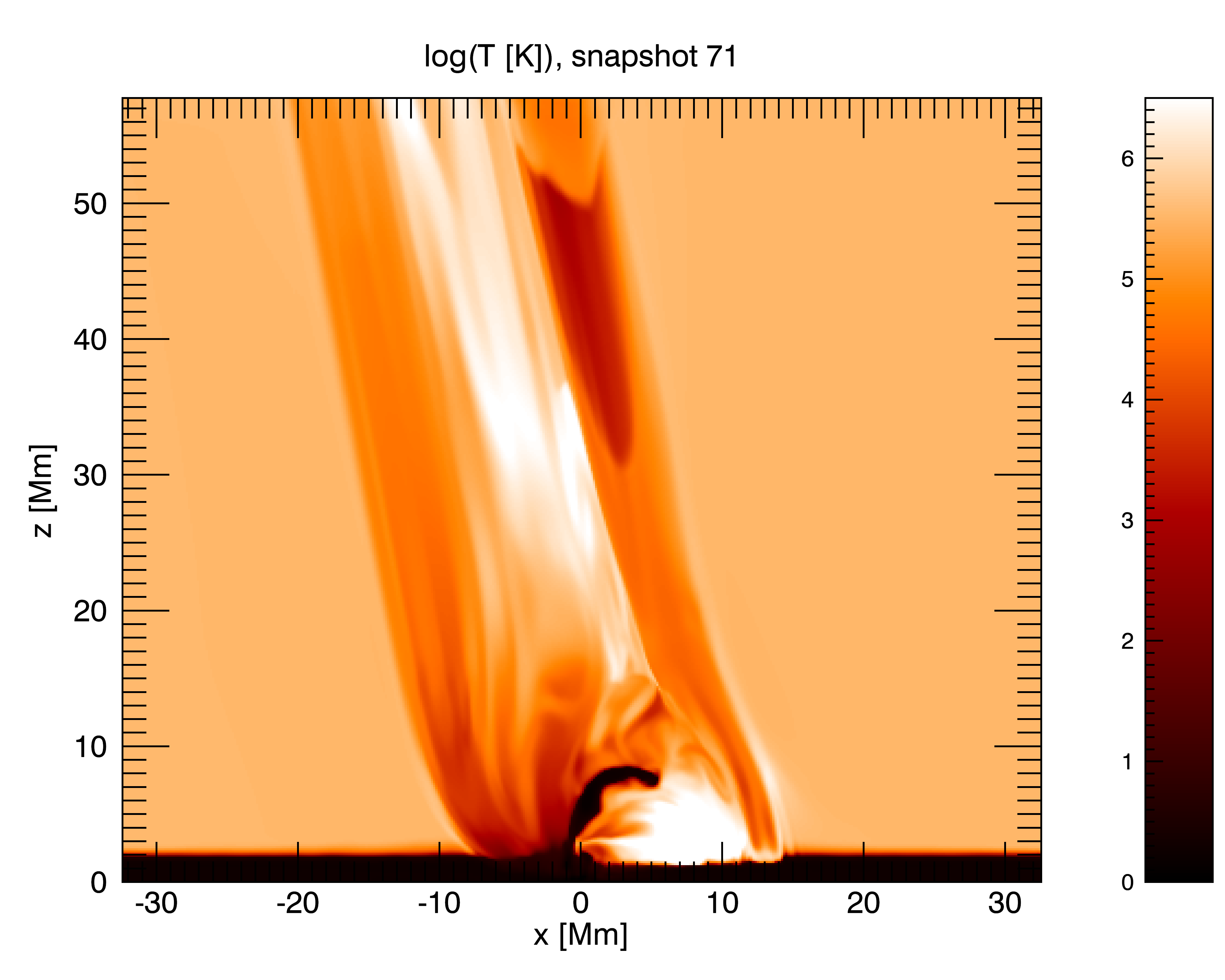}%
\includegraphics[width=0.24\textwidth]{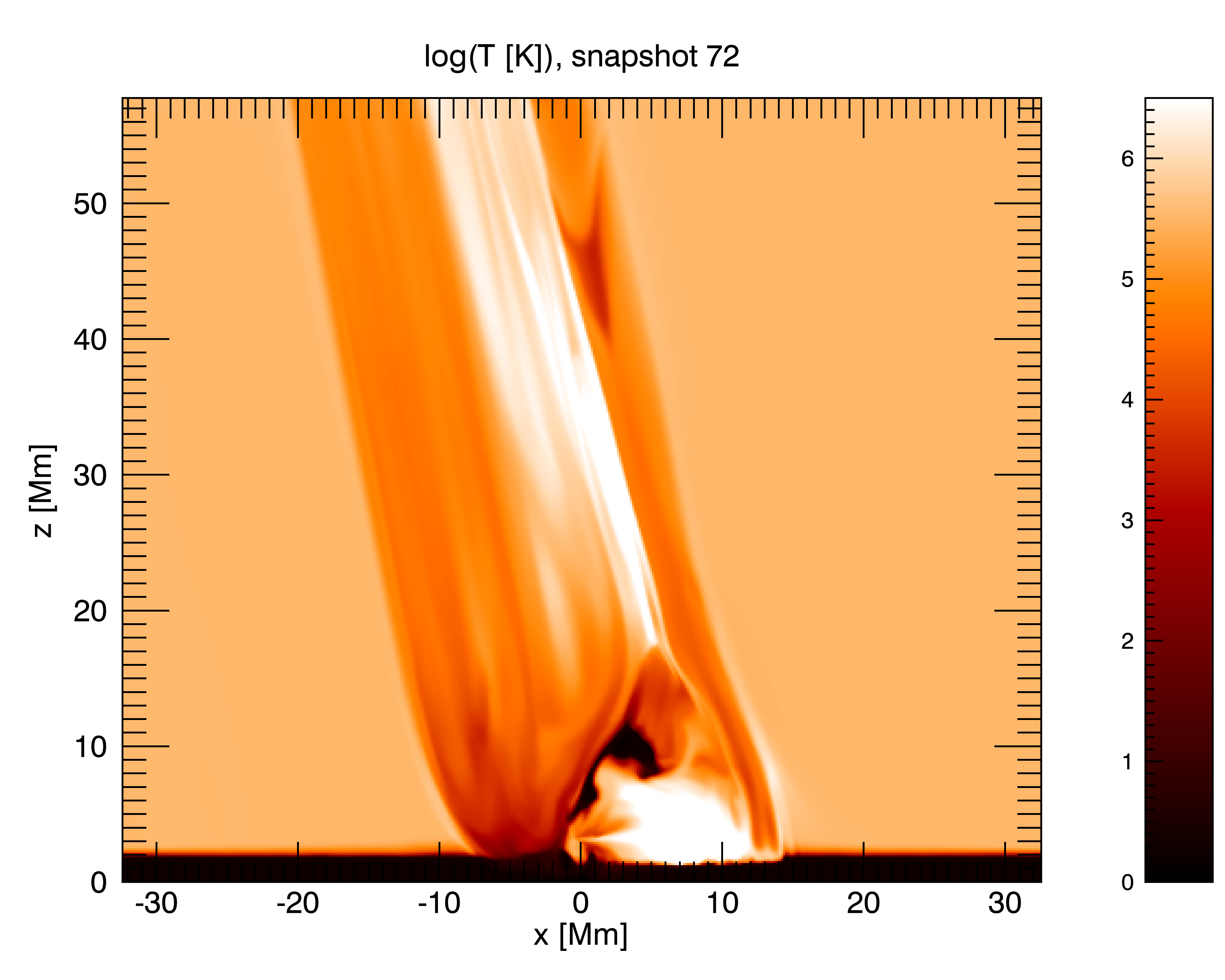}%
\includegraphics[width=0.24\textwidth]{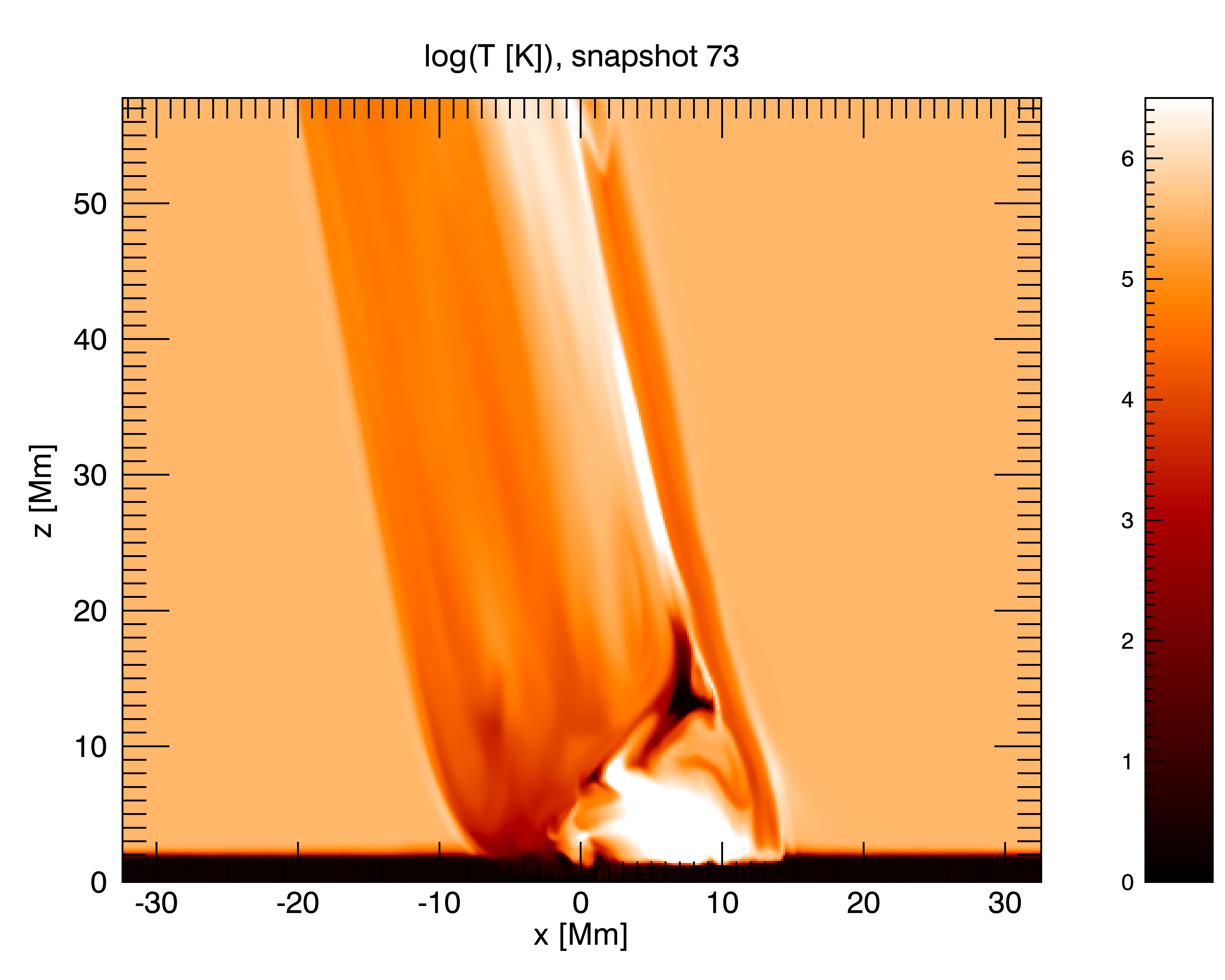}%
\includegraphics[width=0.24\textwidth]{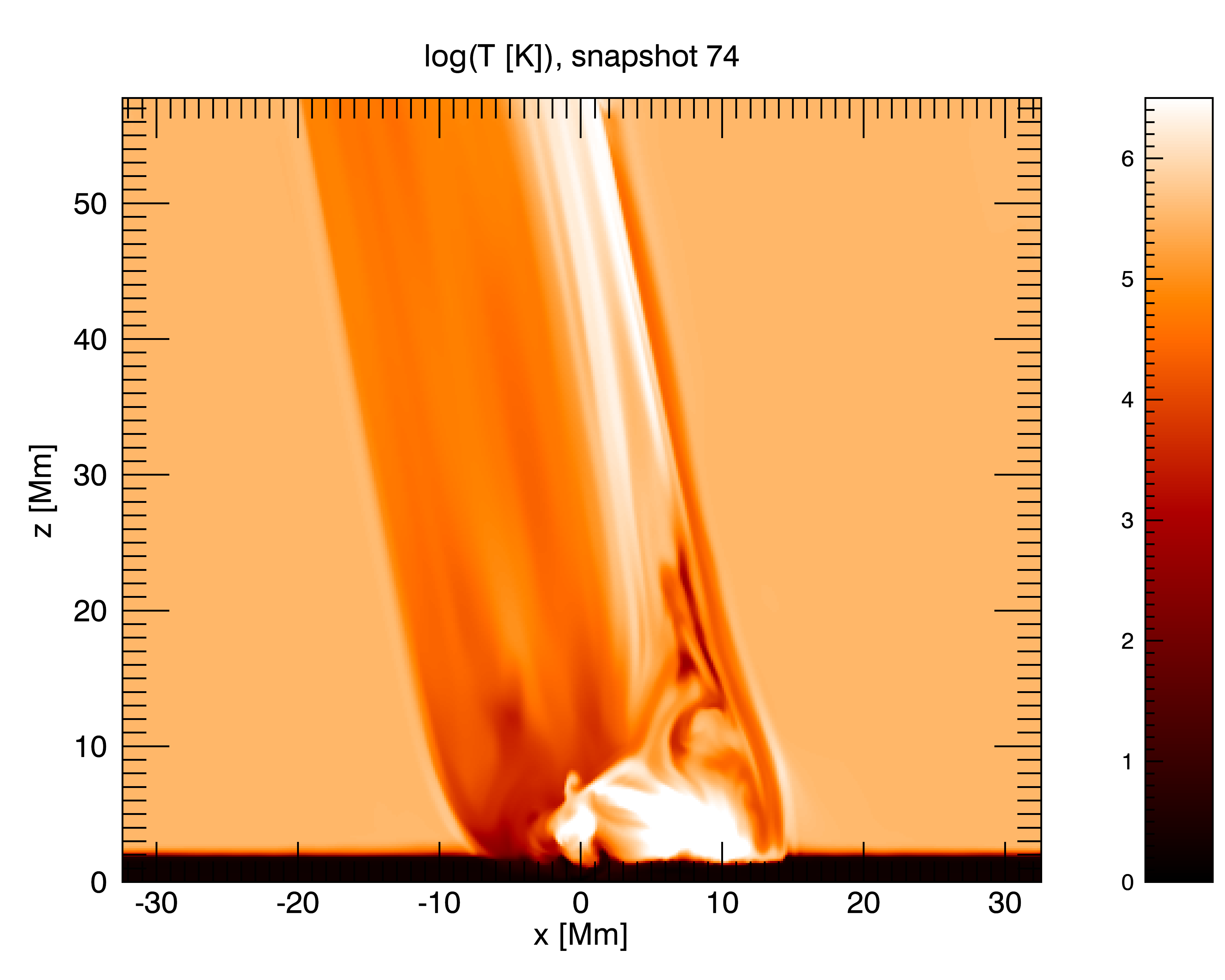}
\caption{Selected snapshots of the temperature morphology (on a logarithmic scale) at the $y=0$ plane. The top row corresponds to the times around the reconnection jet that happens first in the simulation, and the bottom row to the last, and strongest, blowout jet.}
\label{mhdtemp}
\end{figure*}

The evolution of the simulation is quite dynamic, with a number of transient phenomena taking place. More specifically, at snapshots 30-34, which correspond to the times $43\,\mathrm{min}\lesssim t\lesssim 49\,\mathrm{min}$, a reconnection jet is produced when the emerging flux tube meets the ambient magnetic field. Later, three more jets are produced at times $t\sim 70\,\mathrm{min}$, $t\sim 83\,\mathrm{min}$, and $t\sim 106\,\mathrm{min}$, which can all be characterised as blowout jets. We show in Fig.~\ref{mhdtemp} two of the jets, as these are determined from the temperature images in the $y=0$ plane that is perpendicular to the flux tube axis: the reconnection jet, and the last blowout jet, which is the strongest. It should be noted that the exact determination of the jet production times is not possible, and so in the following we indicate a wider time interval around them. Apart from these four events, many other smaller events take place during the simulation.

\section{Methodology}
\label{sect:method}

In this section, we describe the various physical quantities that we are going to study and their computation method.

\subsection{Energies}

The energy of a magnetic field, $\mathbf{B}$, that occupies the volume, $V$, is given by
\begin{equation}
E=\frac{1}{8\pi}\int_V \mathbf{B}^2\,{\rm d}V.
\label{nrg1}
\end{equation}
Another magnetic field of interest is the potential, or current-free field, $\mathbf{B}_\mathrm{p}$, whose normal components have the same distribution as those of $\mathbf{B}$ on the boundary of the volume, $\partial V$. This condition can be written as
\begin{equation}
\left. \hat{n}\cdot \mathbf{B} \right|_{\partial V}=\left. \hat{n}\cdot \mathbf{B}_\mathrm{p} \right|_{\partial V},
\label{helc}
\end{equation}
with $\hat{n}$ denoting the outward-pointing unit normal on $\partial V$.

The energy of the potential field, $E_\mathrm{p}$, was computed from Eq.~(\ref{nrg1}) with $\mathbf{B}_\mathrm{p}$ in place of $\mathbf{B}$. The difference between the two energies defines the free energy, $E_\mathrm{j}=E-E_\mathrm{p}$. Alternatively, free energy can be defined from Eq.~(\ref{nrg1}), with the current-carrying magnetic field $(\mathbf{B}-\mathbf{B}_\mathrm{p})$ in place of $\mathbf{B}$. The two definitions are equivalent when $\mathbf{B}$ is numerically solenoidal enough.

\subsection{Relative helicities}

Relative magnetic helicity \citep{BergerF84,fa85} is defined as
\begin{equation}
H_\mathrm{r}=\int_V (\mathbf{A}+\mathbf{A}_\mathrm{p})\cdot (\mathbf{B}-\mathbf{B}_\mathrm{p})\,{\rm d}V,
\label{helr}
\end{equation}
where $\mathbf{A}$, $\mathbf{A}_\mathrm{p}$ are the vector potentials that correspond to the two magnetic fields. Relative helicity is independent of the gauges of the vector potentials as long as the condition of Eq.~(\ref{helc}) holds.

Relative helicity can be split into two gauge-independent components \citep{berger99}: the current-carrying helicity,
\begin{equation}
H_\mathrm{j}=\int_V (\mathbf{A}-\mathbf{A}_\mathrm{p})\cdot (\mathbf{B}-\mathbf{B}_\mathrm{p})\,{\rm d}V,
\label{helj}
\end{equation}
and the volume-threading helicity,
\begin{equation}
H_\mathrm{pj}=2\int_V \mathbf{A}_\mathrm{p}\cdot (\mathbf{B}-\mathbf{B}_\mathrm{p})\,{\rm d}V.
\label{helpj}
\end{equation}
It is easy to check that summing the two components recovers the relative helicity. 
During the evolution of a magnetic system, the two terms fluctuate but always sum up to the value of the relative helicity. The rate of helicity that is exchanged between the two components is given by
\begin{equation}
\frac{{\rm d} H_\mathrm{j}}{{\rm d} t}=-2 \int_V (\mathbf{v}\times\mathbf{B})\cdot \mathbf{B}_\mathrm{p}\,{\rm d}V,
\label{dhdt}
\end{equation}
where $\mathbf{v}$ is the plasma velocity \citep{linan18}. It should be noted that this quantity is gauge-independent and that its sign corresponds to the transfer of helicity from $H_\mathrm{j}$ to $H_\mathrm{pj}$; the inverse transfer has the opposite sign.

\subsection{Relative field line helicities}

Relative field line helicity, $h_r$, is a proxy for the density of relative helicity \citep{yeates18,moraitis19}. It is a generalisation of the plain field line helicity (FLH) \citep{berger88} and, like FLH, it depends on the gauges of the vector potentials. Relative helicity can be written with the help of RFLH as a surface, and not as a volume integral; namely,
\begin{equation}
H_\mathrm{r,fl}=\oint_{\partial V} h_\mathrm{r}\,{\rm d}\Phi,
\label{flhhel}
\end{equation}
where ${\rm d}\Phi=\left| \hat{n}\cdot\mathbf{B} \right|\,{\rm d}S$ is the elementary magnetic flux on the boundary and ${\rm d}S$ the respective area element. By defining the RFLH operator,
\begin{equation}
h(\mathbf{A})=\int_{\alpha_+}^{\alpha_-}\,\mathbf{A}\cdot {\rm d}\bm{l} - \frac{1}{2}\left( \int_{\alpha_{+}}^{\alpha_{p-}}\,\mathbf{A} \cdot {\rm d}\bm{l}_\mathrm{p}+\int_{\alpha_{p+}}^{\alpha_{-}}\,\mathbf{A} \cdot {\rm d}\bm{l}_\mathrm{p} \right),
\label{flhdefgen}
\end{equation}
we can simply write RFLH as
\begin{equation}
h_\mathrm{r}=h(\mathbf{A}+\mathbf{A}_\mathrm{p}).
\label{flhhr}
\end{equation}
In Eq.~(\ref{flhdefgen}), $\alpha_+$, $\alpha_-$, ${\rm d}\bm{l}$ denote the positive footpoint, the negative footpoint, and the elementary length along the field lines of $\mathbf{B}$, respectively, while $\alpha_{p+}$, $\alpha_{p-}$, ${\rm d}\bm{l}_\mathrm{p}$ denote those of $\mathbf{B}_\mathrm{p}$.

The current-carrying component of relative helicity can be expressed through a current-carrying RFLH, $h_\mathrm{j}$, as
\begin{equation}
H_\mathrm{j,fl}=\oint_{\partial V} h_\mathrm{j}\,{\rm d}\Phi,
\label{flhhelj}
\end{equation}
similarly to Eq.~(\ref{flhhel}). The corresponding RFLH is 
\begin{equation}
h_\mathrm{j}=h(\mathbf{A}-\mathbf{A}_\mathrm{p}).
\label{flhhj}
\end{equation}
From the difference between the two RFLHs, we can also define the RFLH of the volume-threading helicity, as
\begin{equation}
h_\mathrm{pj}=h_\mathrm{r}-h_\mathrm{j}=h(2\mathbf{A}_\mathrm{p}).
\end{equation}

\subsection{Numerical computation of the various quantities}

The computation of the potential field satisfying the condition of Eq.~(\ref{helc}) was performed with the numerical solution of a 3D Laplace equation \citep{moraitis14}. In the computation of the vector potentials from the respective magnetic fields, it is assumed that they satisfy the \citet[][DV]{devore00} gauge, as this was specialised in \citet{val12}. More specifically, $\mathbf{A}$ is taken in the simple DV gauge and $\mathbf{A}_\mathrm{p}$ in the Coulomb DV gauge (DVS and DVC in the notation of \citet{moraitis18}). The field line integrations required for computing the RFLHs were performed in the manner described in \citet{moraitis19}. Similar to that work, the footpoints in the RFLH computations are restricted on the `photospheric' boundary, the plane $z=0$. We finally note that all integrals were computed using the trapezoidal rule.

\section{Results}
\label{sect:results}

In this section, we examine the evolution of the various quantities discussed in Sect.~\ref{sect:method} in three cases: when the whole simulation volume is considered, when different subvolumes of the total volume are considered, and when a 2D region on the photosphere is considered. We start with the first case.

\subsection{Consideration of the whole volume}
\label{sect:results1}

\begin{figure}[ht]
\centering
\includegraphics[width=0.46\textwidth]{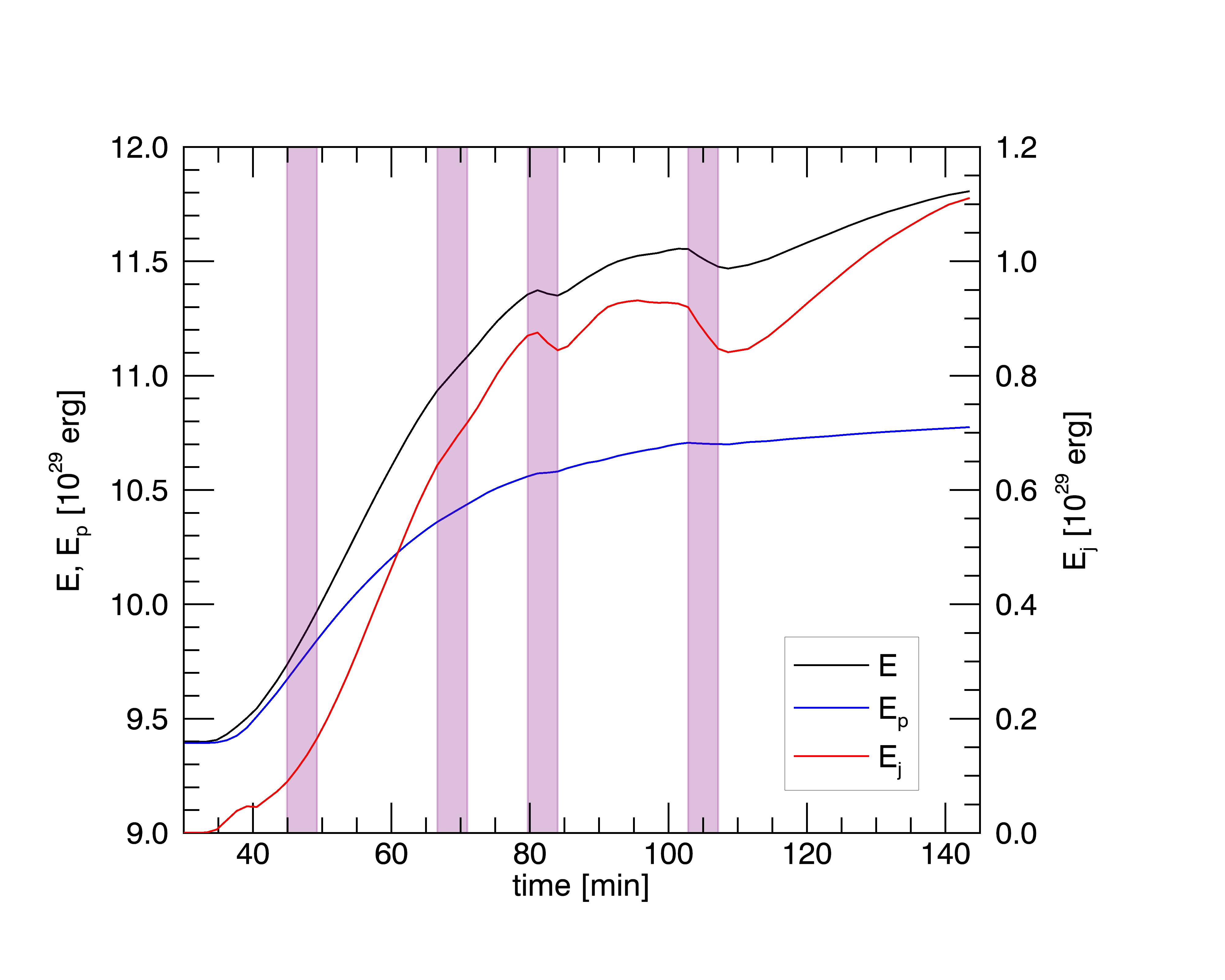}
\caption{Evolution of the total, potential, and free magnetic energies. The vertical purple stripes represent the time intervals around the four identified jets.}
\label{nrgsevol}
\end{figure}

The evolution of the various energies is shown in Fig.~\ref{nrgsevol}. The energy of the magnetic field (black curve) is in general an increasing function of time, which exhibits small decreases during the last two jets. The potential energy (blue curve) shows a mild increase that after $t\sim 80,\mathrm{min}$ becomes much slower. The free energy (red curve) has two increasing periods, one until the third jet, and another one after the fourth jet, while in between it fluctuates a lot due to the production of the blowout jets. The decreases experienced by the free energy during the last two jets are more pronounced compared to those of the total energy. In all plots, the evolution after $t=30$~min, or snapshot number 20, is shown, since before that all of the quantities are practically constant. Something worth noticing is that the total and potential energies start from non-zero values due to the presence of the ambient magnetic field. This is not the case for the free energy, which initially is zero.

\begin{figure}[ht]
\centering
\includegraphics[width=0.46\textwidth]{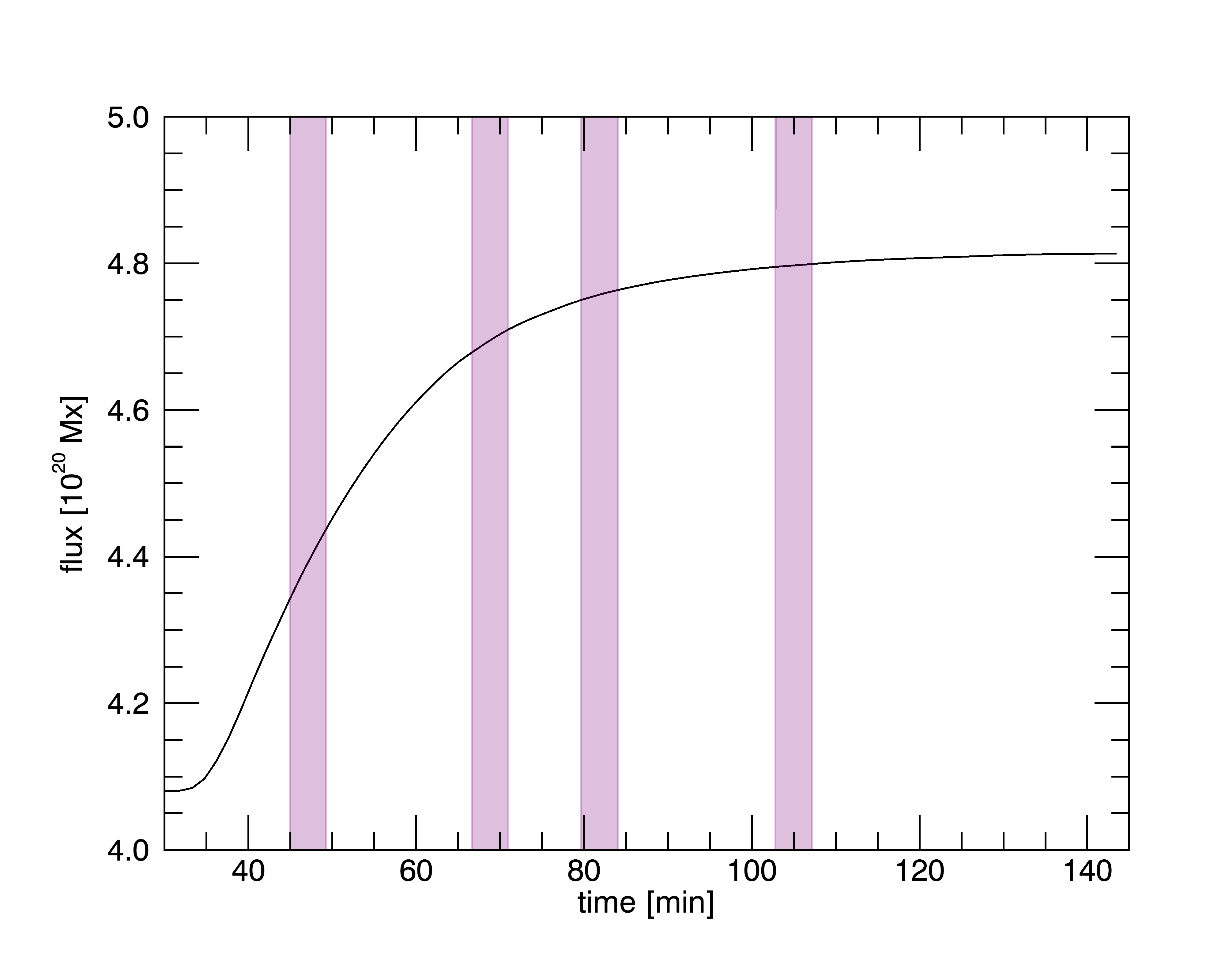}
\caption{Evolution of the unsigned magnetic flux. The vertical purple stripes represent the time intervals around the four identified jets.}
\label{flxevol}
\end{figure}

The reason for not getting significant jet-related changes during the first two jets in the energy patterns can be deduced from the evolution of the unsigned magnetic flux, which is shown in Fig.~\ref{flxevol}. We notice that during the first two jets flux is still emerging at a high rate and this leads to a similarly steep increase in the energy curves and the suppression of any finer details. The flattening of the flux evolution during the last two jets allows the jet-related changes to stand out more clearly.
 
\begin{figure}[ht]
\centering
\includegraphics[width=0.46\textwidth]{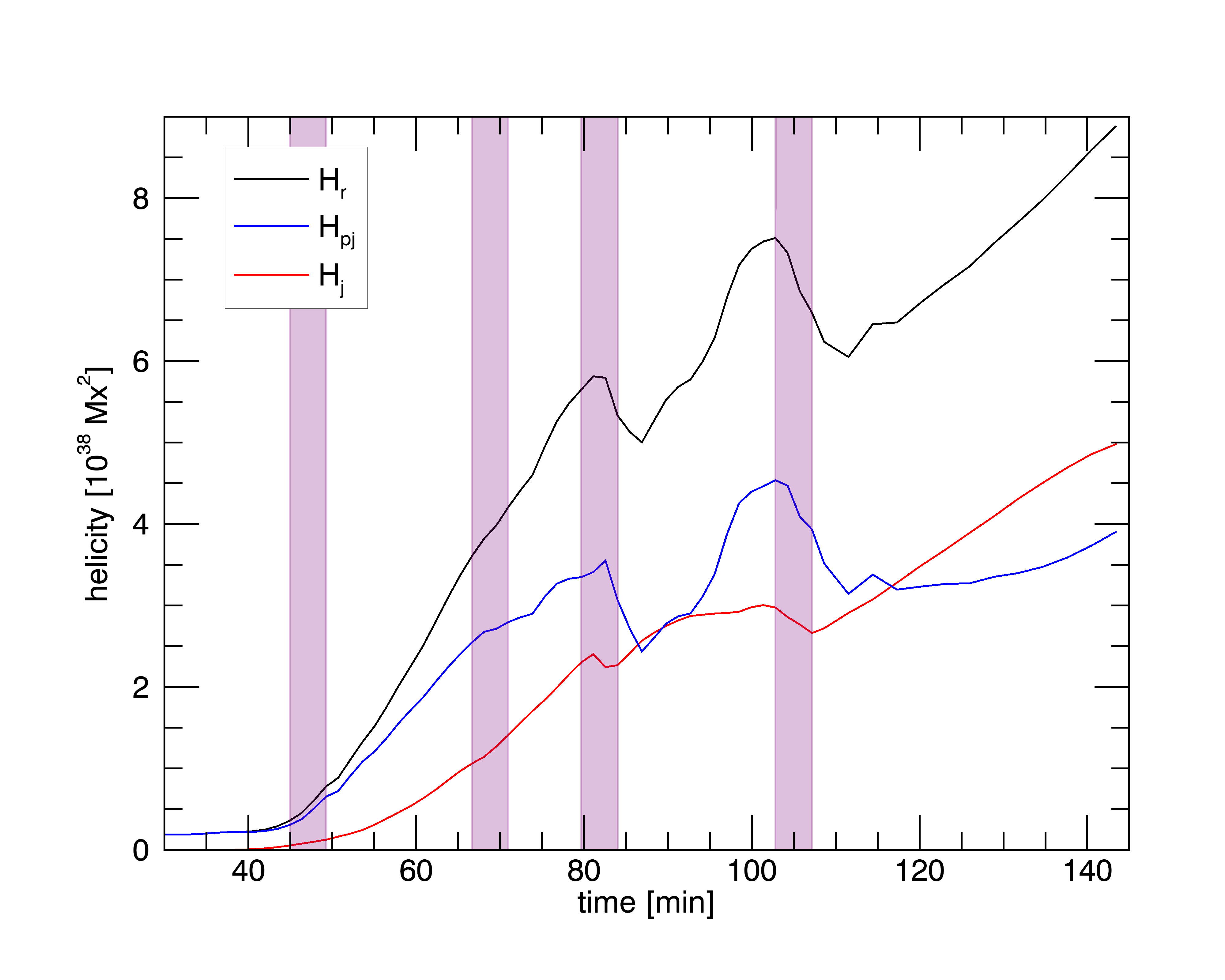}
\caption{Evolution of the relative, current-carrying, and volume-threading helicities. The vertical purple stripes represent the time intervals around the four identified jets.}
\label{helsevol}
\end{figure}

The evolution of the various helicities is shown in Fig.~\ref{helsevol}. The relative helicity (black curve) exhibits an overall increasing pattern that is interrupted by large decreases during the last two events. The relative helicity also experiences small changes during the first two jets, when the rate of its increase weakens. The volume-threading helicity (blue curve) follows the same pattern as relative helicity, with the difference that it flattens after the large blowout jet and does not continue to increase as relative helicity does. The current-carrying helicity (red curve) does not show important changes during the first two events but decreases during the last two jets like the other helicities, although to a lesser degree. This seems counter-intuitive as the non-potential $H_\mathrm{j}$ should show more evident changes during jet eruptions compared to $H_\mathrm{pj}$. It can be explained however by the presence of the ambient magnetic field, which amplifies the volume-threading component, the mutual helicity of the ambient and emerging fields, and its variations. The behaviour of the current-carrying helicity is also different between the third and fourth jets, when it increases much more mildly than the other helicities. As also happens for energies, the relative and volume-threading helicities start from non-zero values owing to the presence of the ambient magnetic field. The current-carrying helicity is initially zero, since in the beginning of the simulation all the current-carrying magnetic field is located in the not-yet-emerged flux tube. Finally, the peaks of the current-carrying helicity in the last two events, which are the largest, occur a little earlier than those of $H_\mathrm{pj}$ and $H_\mathrm{r}$.

\begin{figure}[ht]
\centering
\includegraphics[width=0.46\textwidth]{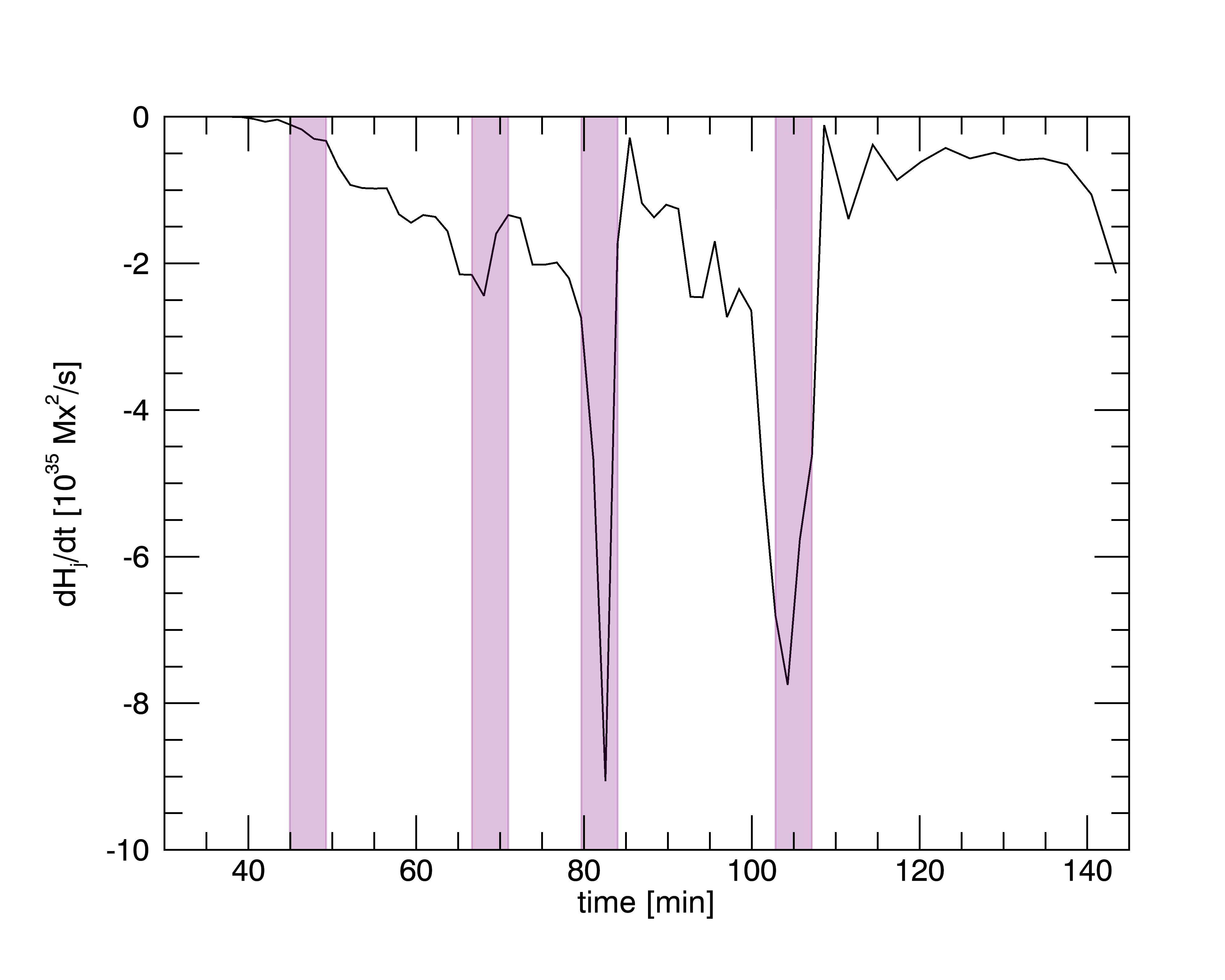}
\caption{Evolution of the $H_\mathrm{j}$ to $H_\mathrm{pj}$ helicity transfer term. The vertical purple stripes represent the time intervals around the four identified jets.}
\label{dhjdt}
\end{figure}

The dynamics of the two relative helicity components can be seen in more detail in Fig.~\ref{dhjdt}, in which the $H_\mathrm{j}$ to $H_\mathrm{pj}$ transfer term, which was introduced in the analysis of \citet{linan18}, is depicted. We note that the three major peaks of this transfer term coincide with the times of the three largest events, the blowout jets. During these time intervals, the transfer term increases and then relaxes abruptly, which means that there is an increased transformation of $H_\mathrm{j}$ to $H_\mathrm{pj}$ then. This can also be explained by Fig.~\ref{helsevol}, in which $H_\mathrm{j}$ peaks earlier than $H_\mathrm{pj}$, and so, in between the peaks of the two helicities, the former decreases while the latter increases.

\begin{figure}[ht]
\centering
\includegraphics[width=0.46\textwidth]{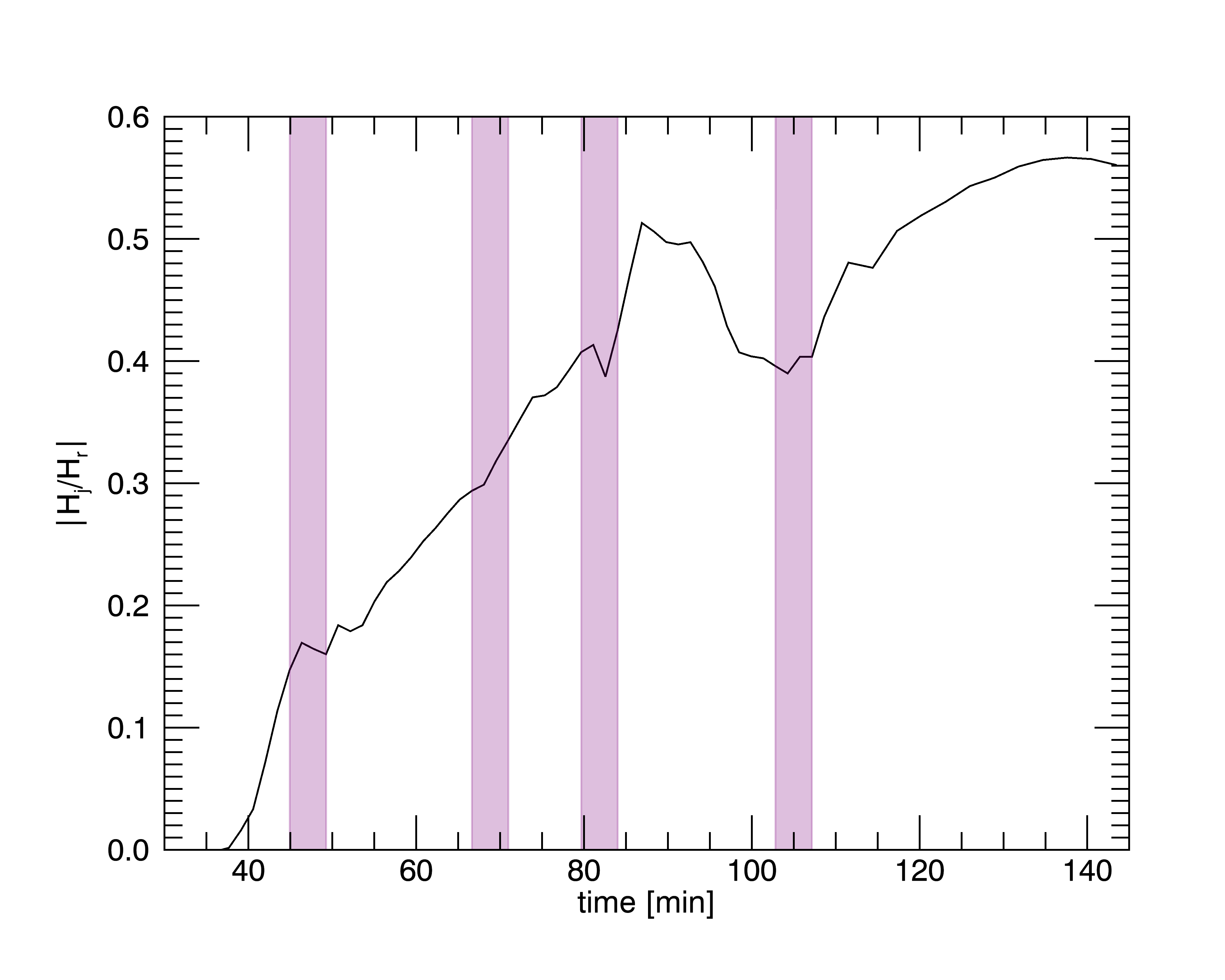}
\caption{Evolution of the helicity eruptivity index. The vertical purple stripes represent the time intervals around the four identified jets.}
\label{hjhrevol}
\end{figure}

The incoherent behaviour of $H_\mathrm{j}$ and $H_\mathrm{pj}$ during the simulation leads to an irregular behaviour of the eruptivity index; that is, of the ratio $|H_\mathrm{j}|/|H_\mathrm{r}|$, as is shown in Fig.~\ref{hjhrevol}. We note that the general trend of the eruptivity index is increasing more or less until $t\sim85,\mathrm{min}$. It then drops until the large blowout jet and then increases again but more slowly. The eruptivity index experiences changes during all jets, which are mostly decreases. During the reconnection jet, it decreases a bit, while also exhibiting some jiggling; during the second jet, it shows a small break-like decrease; during the third jet, it drops more intensely but then rises even more; and in the large blowout jet, it shows a small decrease followed by an increase. The eruptivity index also exhibits various changes outside of the intervals of jet activity, most notably with the peak around $t\sim85\,\mathrm{min}$.

\subsection{Consideration of three subvolumes}
\label{sect:results2}

We next consider three separate subvolumes of the whole volume and examine the evolution of all of the quantities in them. We first restricted the horizontal span of the subvolumes to $-20\,\mathrm{Mm}<x,y<20\,\mathrm{Mm}$, as the magnetic field is very weak outside this region. We also truncated the height of the total volume to $z<30\,\mathrm{Mm}$. The remaining volume was split into three unequal subvolumes with height ranges $1\,\mathrm{Mm}<z<6\,\mathrm{Mm}$, $6\,\mathrm{Mm}<z<11\,\mathrm{Mm}$, and $11\,\mathrm{Mm}<z<30\,\mathrm{Mm}$. The choice of $6\,\mathrm{Mm}$ and $11\,\mathrm{Mm}$ for the subvolumes' height limits follows from the temperature morphology of Fig.~\ref{mhdtemp} and should be considered approximate. The lower subvolume was chosen to depict what happens below the reconnection point of the initial jet, the middle subvolume for where most of the action takes place, and the upper one for the lower-intensity coronal part. Moreover, the first subvolume starts at $z=1\,\mathrm{Mm}$ so as to avoid the turbulent layer at, and slightly above, the photosphere. The two lower subvolumes have almost the same size, while the upper one is $\sim 4$ times their size. The sum of these three volumes is $\sim 20\%$ of the coronal volume studied in Sect.~\ref{sect:results1}. The subvolumes, and their relation with the whole volume, are shown in Fig.~\ref{mhdsetup}.

In each subvolume, we computed the energies and helicities with the methodology described in Sect.~\ref{sect:method}. This requires the computation of a different potential field in each subvolume, which satisfies the appropriate Eq.~(\ref{helc}) for the specific volume. We should stress here that the various helicities are not additive \citep{valori20}; that is, the sum of the helicities of the subvolumes is not equal to the helicity of the whole volume.

\begin{figure*}[ht]
\centering
\includegraphics[width=0.32\textwidth]{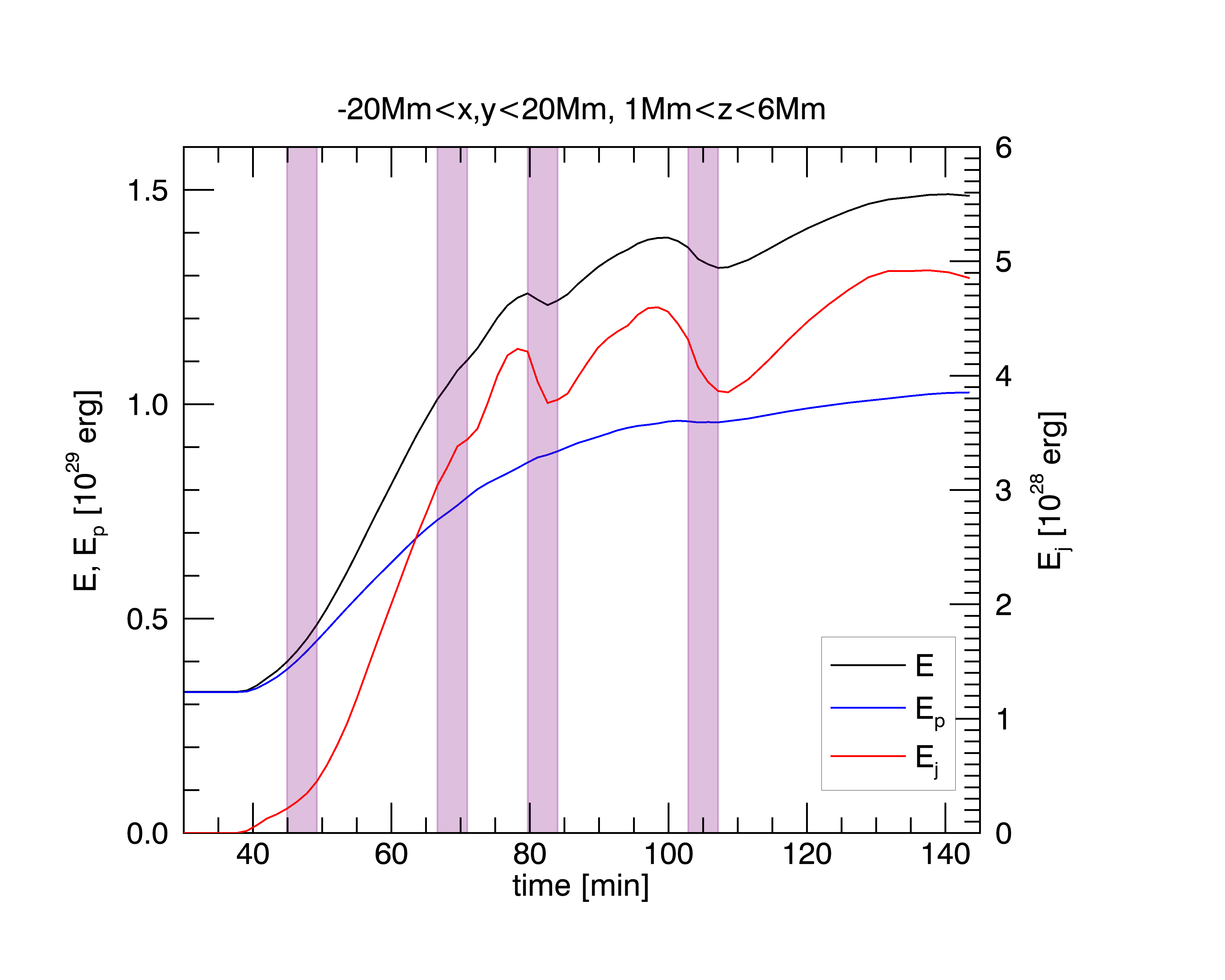}%
\includegraphics[width=0.32\textwidth]{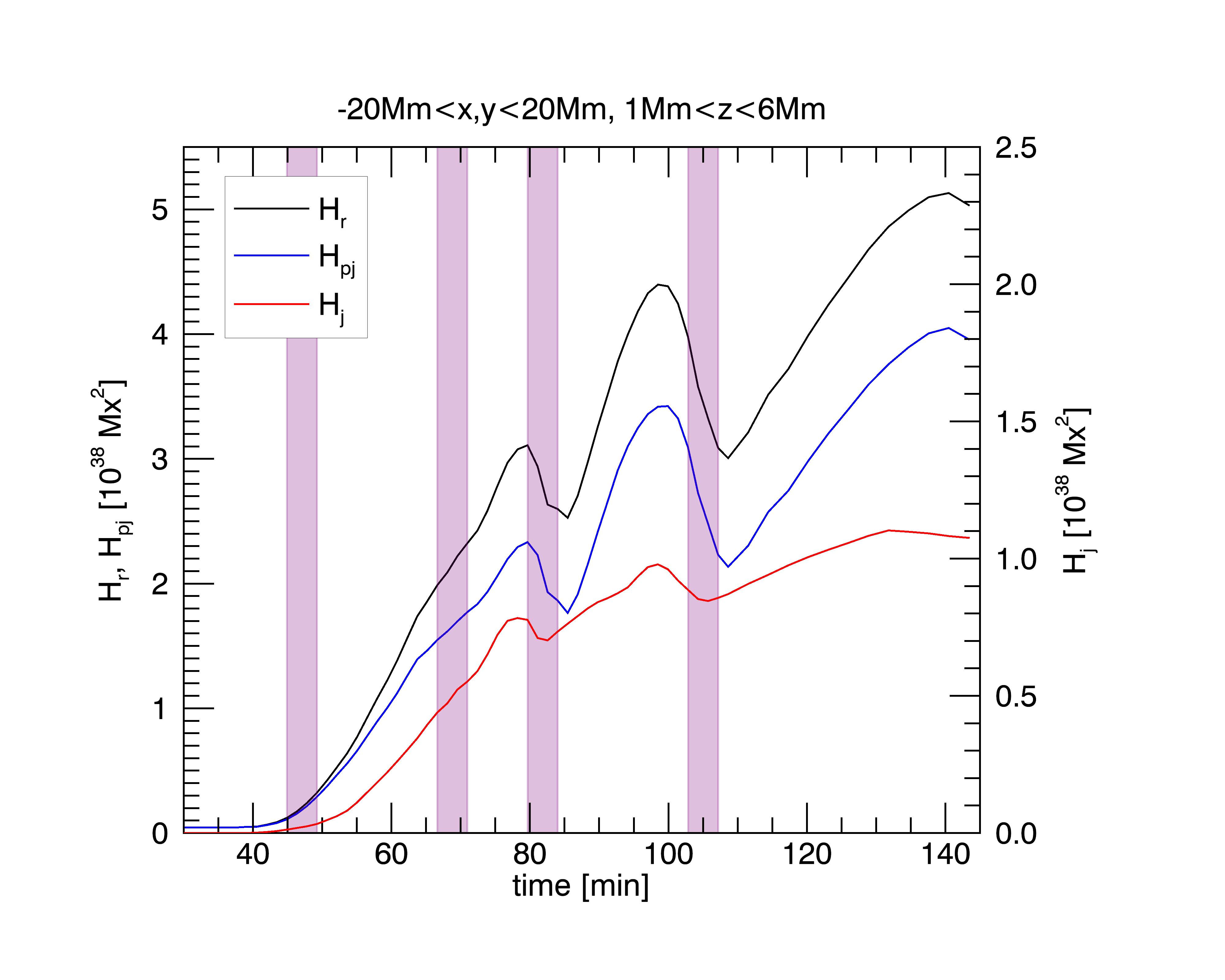}%
\includegraphics[width=0.32\textwidth]{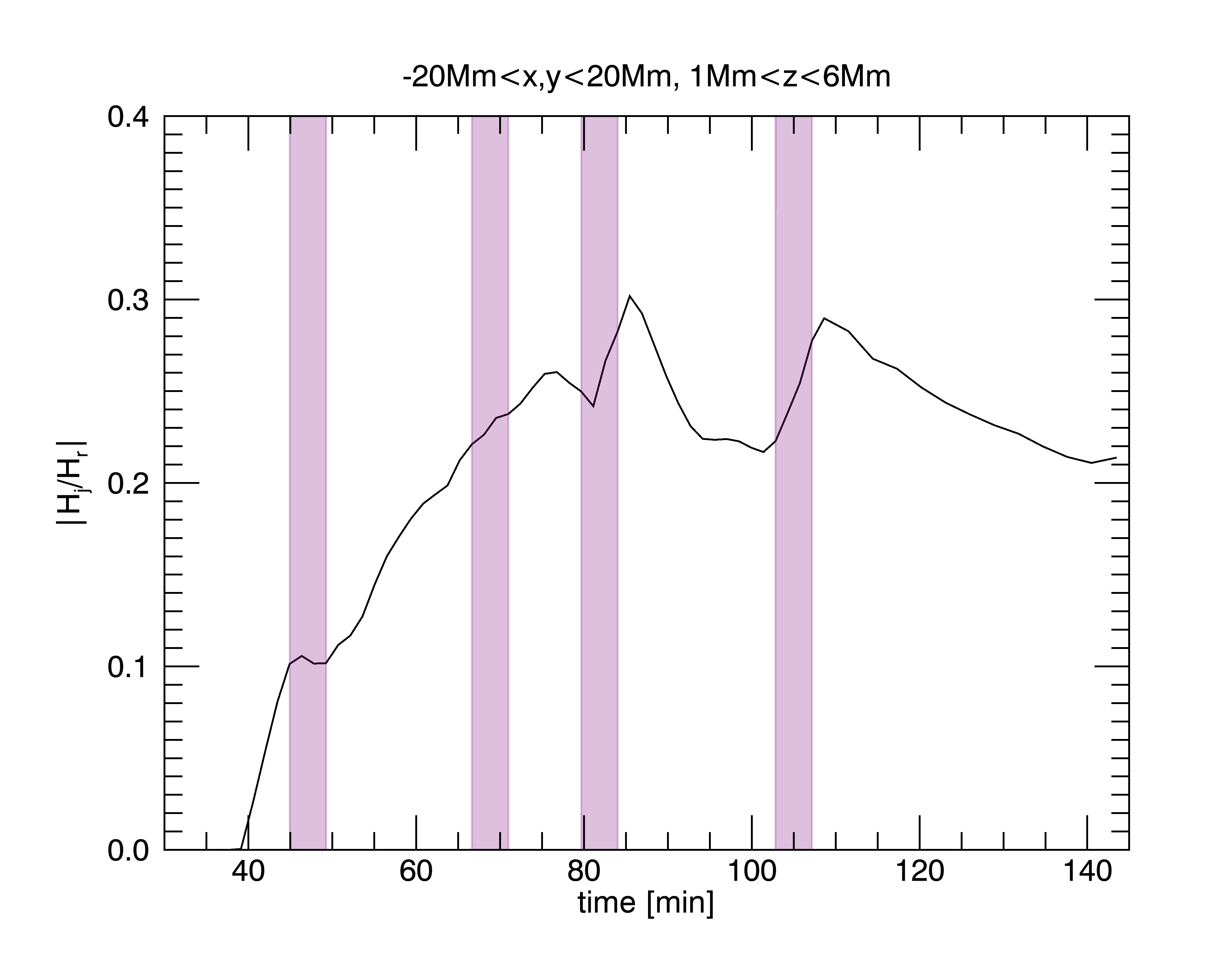}

\includegraphics[width=0.32\textwidth]{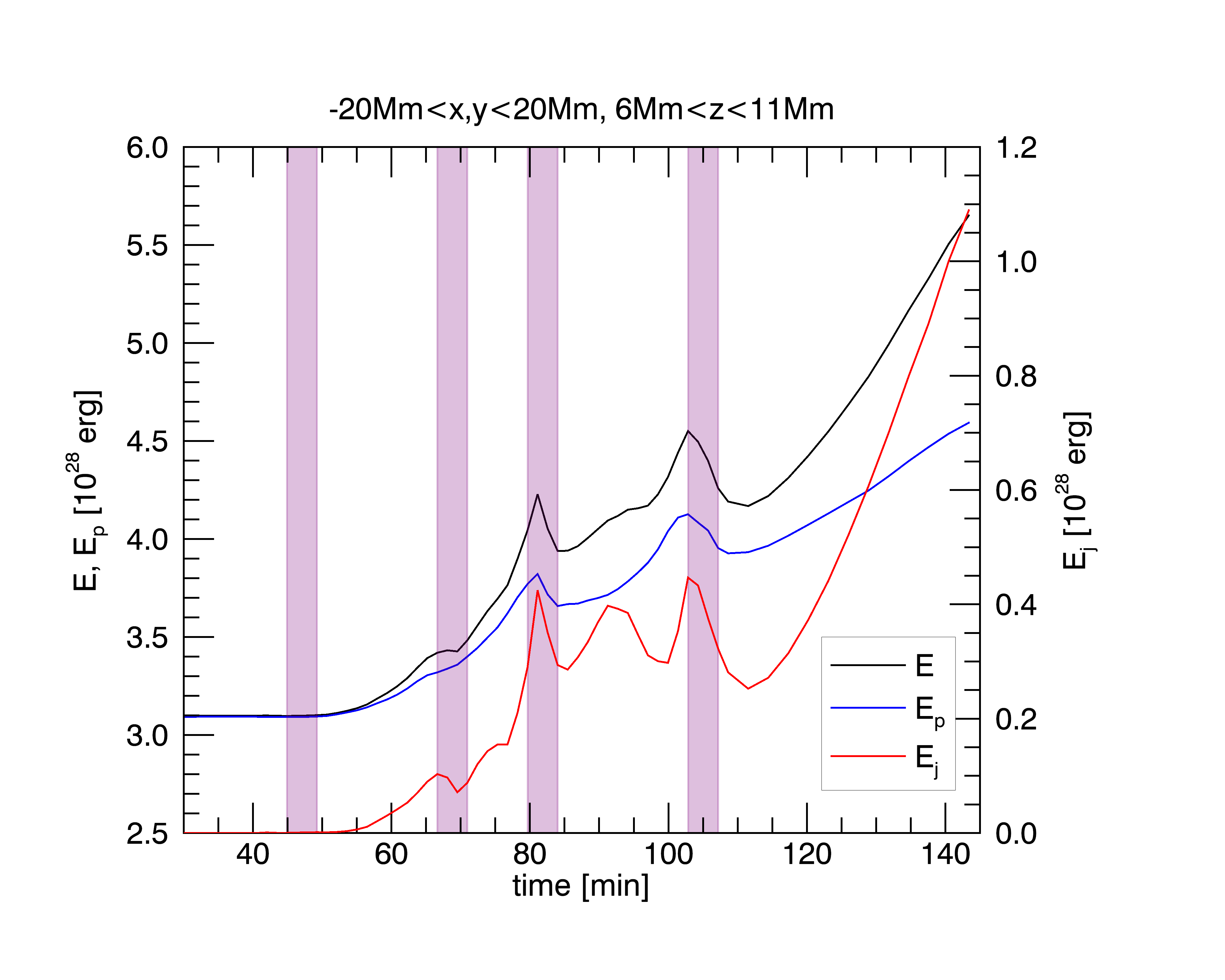}%
\includegraphics[width=0.32\textwidth]{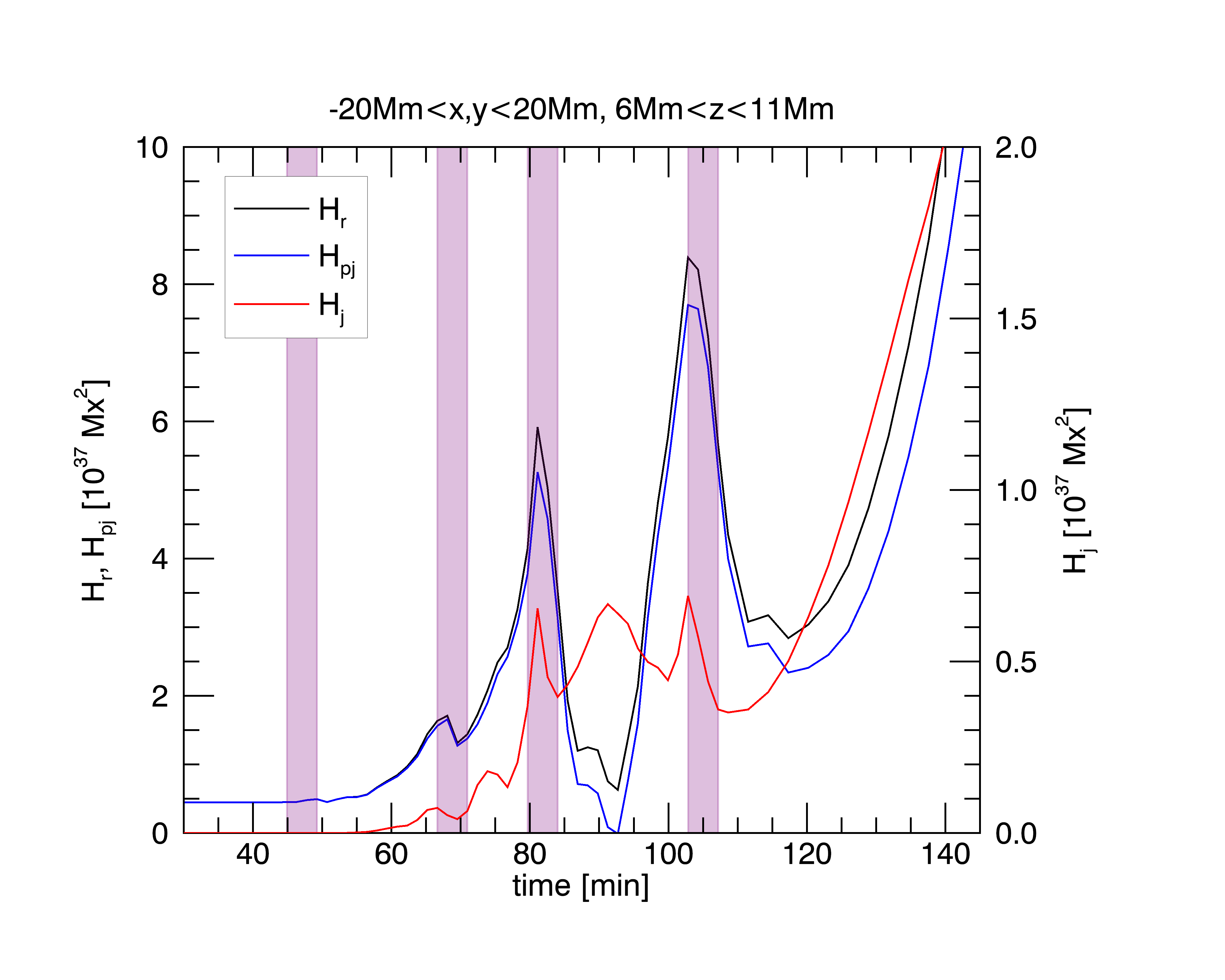}%
\includegraphics[width=0.32\textwidth]{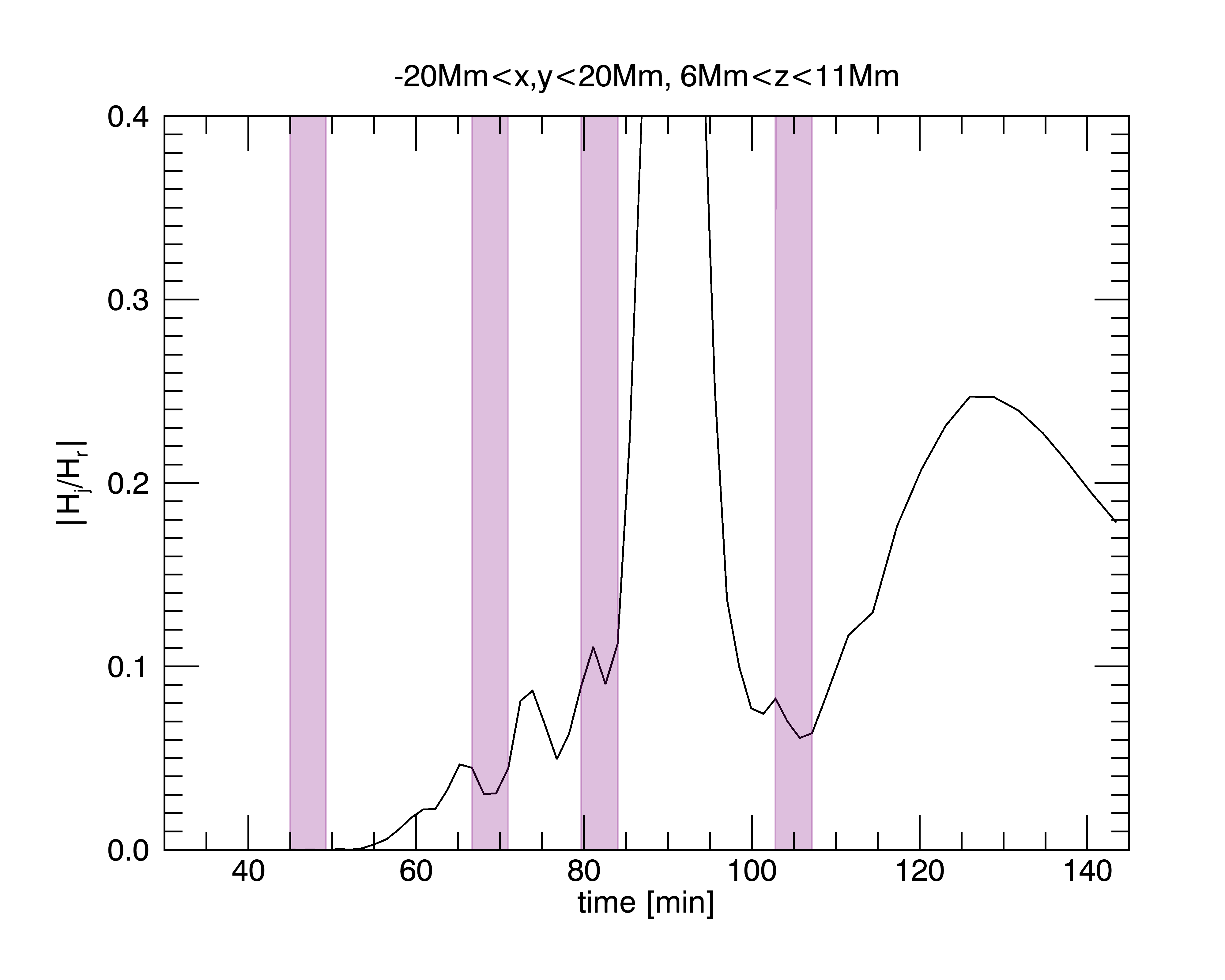}

\includegraphics[width=0.32\textwidth]{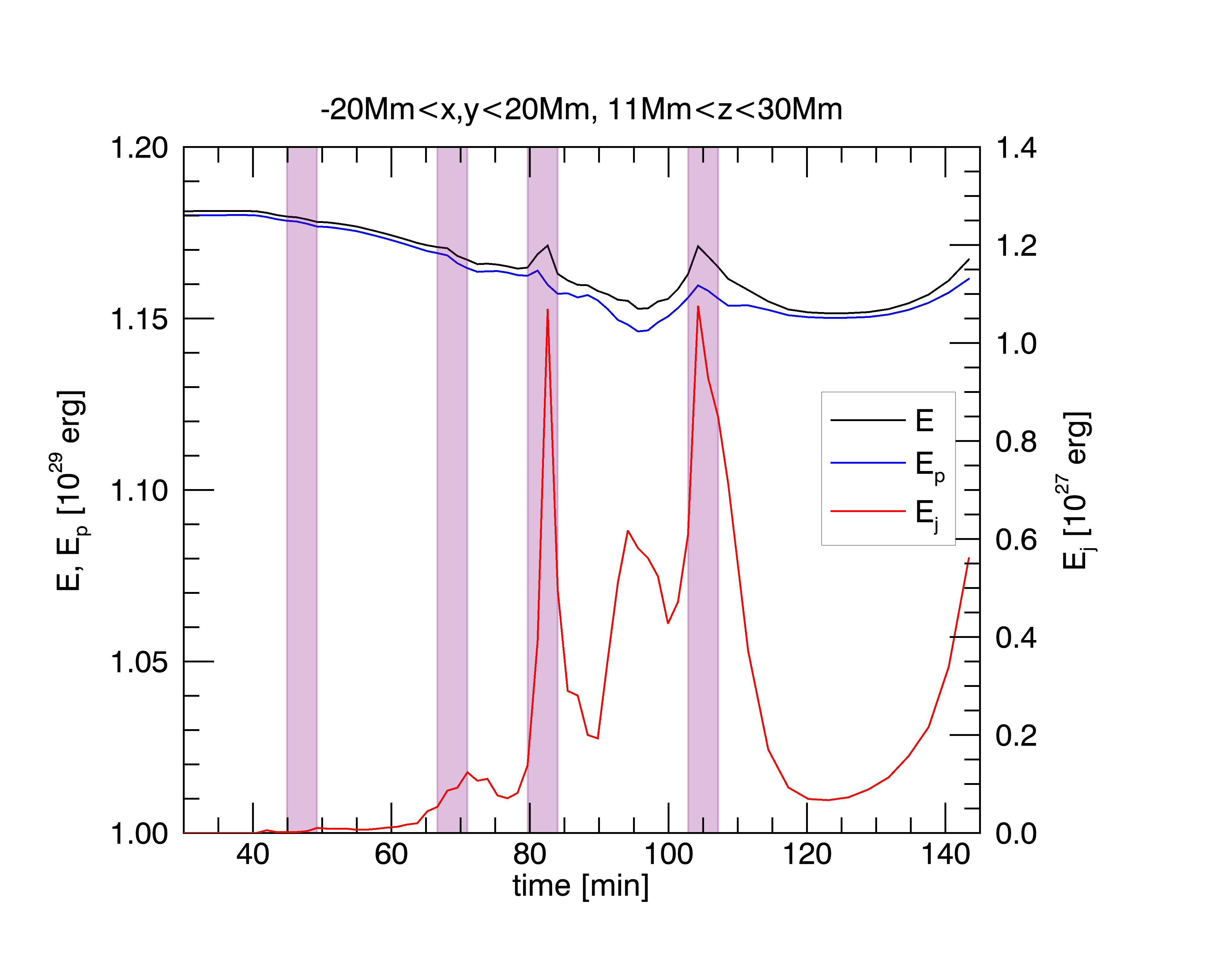}%
\includegraphics[width=0.32\textwidth]{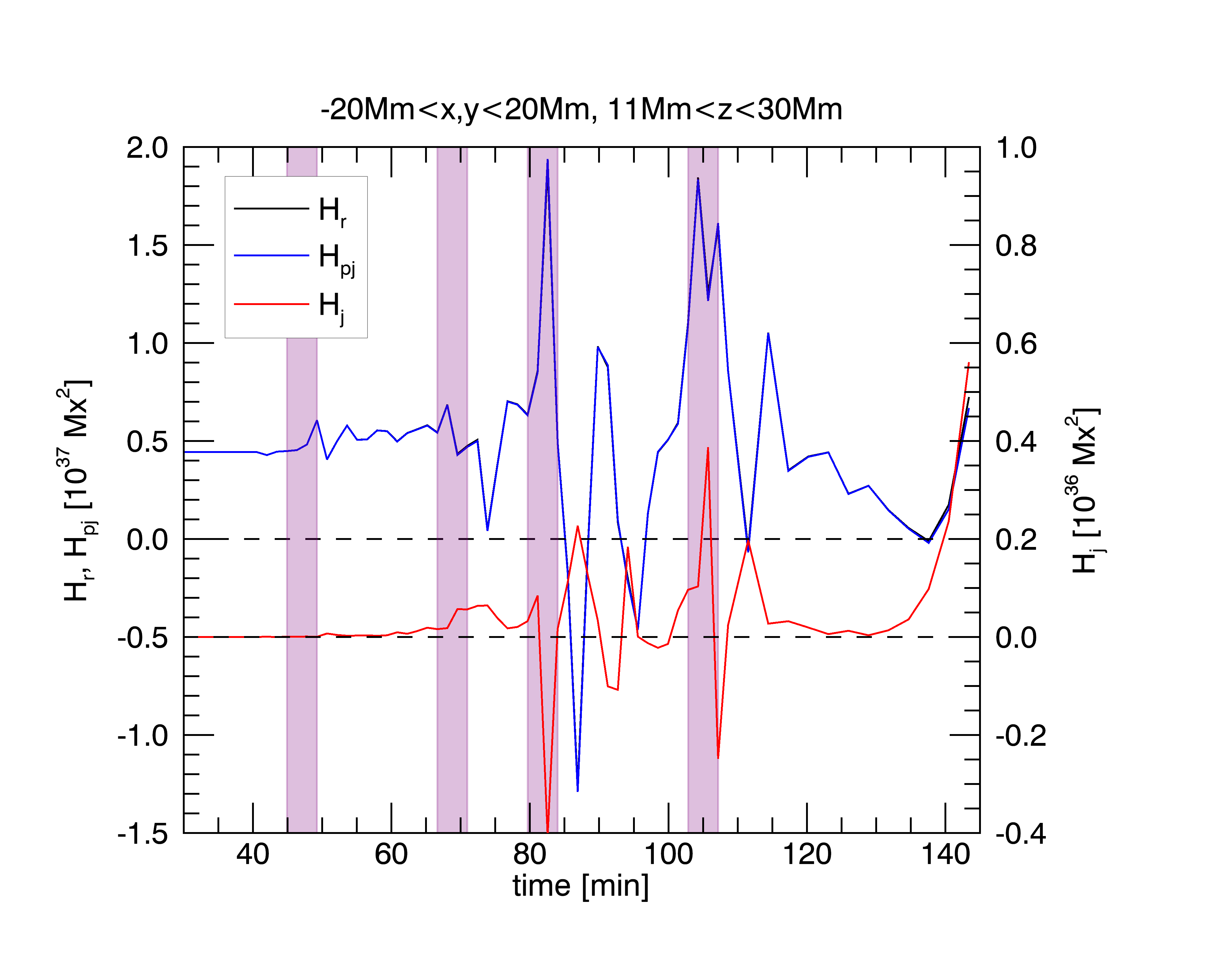}%
\includegraphics[width=0.32\textwidth]{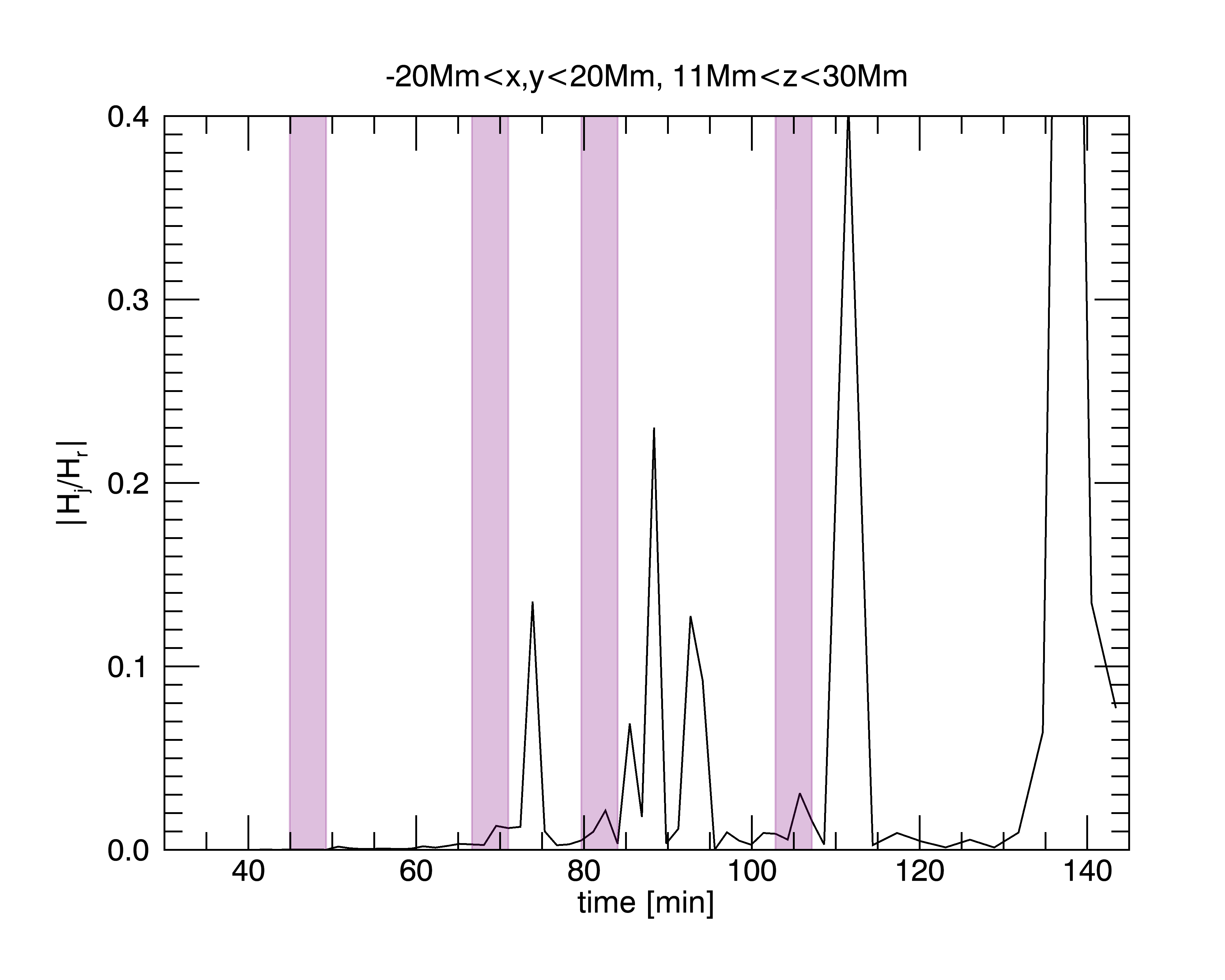}
\caption{Evolution of the energies (left column), the helicities (middle column), and the helicity ratio  (right column), in the lower (top row), middle (middle row), and upper subvolume (bottom row). The vertical purple stripes represent the time intervals around the four identified jets.}
\label{evol_c}
\end{figure*}

The energy and helicity evolution curves for the lower subvolume, at heights $1\,\mathrm{Mm}<z<6\,\mathrm{Mm}$, are shown on the top panel of Fig.~\ref{evol_c}. The energy patterns in this first subvolume (top left panel in Fig.~\ref{evol_c}) are similar to those of Fig.~\ref{nrgsevol}, with the difference that they are a little smoother because of neglecting the lower 1~Mm of the photosphere. Additionally, they are up to an order of magnitude smaller, as was expected from the much smaller volume they come from. The helicity patterns (top middle panel in Fig.~\ref{evol_c}) have an overall similar behaviour to those of Fig.~\ref{helsevol}, with the exception that their peaks occur a little earlier, and also that they start decreasing at the end of the simulation. The helicity ratio pattern (top right panel in Fig.~\ref{evol_c}) has four local maxima, with only the first coincident with the reconnection jet and the rest occurring in between jet activity. 

In the middle subvolume, at heights of $6\,\mathrm{Mm}<z<11\,\mathrm{Mm}$, the energies (middle left panel in Fig.~\ref{evol_c}) are even smaller and have local maxima in all but the reconnection jet. The free energy shows an additional peak between the third and fourth jets, around $t\sim 95\,\mathrm{min}$. The helicity patterns (middle panel in Fig.~\ref{evol_c}) are smaller, different than in Fig.~\ref{helsevol}, and show peaks coincident with the four jets. The current-carrying helicity shows the same additional peak as the free energy. The helicity ratio pattern (middle right panel in Fig.~\ref{evol_c}) is totally different than in Fig.~\ref{hjhrevol}; it exhibits many peaks but only one of them is during the production of a jet, the one during the second blowout jet. 

In the upper subvolume, at heights of $11\,\mathrm{Mm}<z<30\,\mathrm{Mm}$, the total and potential energies (bottom left panel in Fig.~\ref{evol_c}) are mostly decreasing functions of time, with local maxima at the two last jets. The free energy shows a similar pattern as in the other volumes, but it has much smaller values, as it is two orders of magnitude smaller than the total energy. Moreover, it has two sharp peaks during the last two events. The helicity patterns (bottom middle panel in Fig.~\ref{evol_c}) are quite fuzzy and show mixed signs, fluctuating between positive and negative values. During the last two jets, however, $H_\mathrm{j}$ and $H_\mathrm{pj}$ show jet-related changes. The helicity ratio pattern (bottom right panel in Fig.~\ref{evol_c}) is spiky, with various peaks that do not correspond to the jet activity. 

When we look the panels of Fig.~\ref{evol_c} vertically, we can observe a few more things about the height dependence of the various quantities. Focusing first on the energies (left column of Fig.~\ref{evol_c}), we notice that the total energy is a factor of three lower (on average) in the middle volume compared to the other two. The free energy on the other hand decreases by an order of magnitude from the lower to the middle volume, and by a factor of five from the middle to the upper volume. Put simply, the magnetic field becomes more potential as we go higher. The curves for relative and volume-threading helicities decrease by factors of $\sim 5$ as we move to the next higher volume, while the current-carrying helicity by factors of $\sim 15$. This reaffirms the link between the potentiality of the field and the height, and additionally shows that the field gets less helical and twisted upwards. For the helicity ratio, we can only note that its average values decrease with height, and that it exhibits more spikes higher up due to the more frequent close-to-zero values of relative helicity. A final observation is about the relative timings of the peaks of the energy and helicity curves. If we focus on the more pronounced peaks during the last two jets in either of the curves of the left and middle columns of Fig.~\ref{evol_c}, we see a slight shift to the right as we go to higher volumes. This could be a signature of the time it needs for the disturbances to propagate from one volume to the other.

\subsection{Evolution of polarity inversion line helicities}
\label{sect:results3}

Apart from the volumes examined in the previous sections, a 2D region of interest constitutes the PIL of the magnetic field on the photosphere. In a recent work, \citet{moraitis24} found that the part of the relative helicity that is contained in the region around the PIL can be used to indicate solar eruptive behaviour. We now examine whether this result is relevant for a simulated solar AR as well.

As in that work, we used the method of \citet{schrijver07} to determine the region of interest around the PIL. More specifically, we considered a dilation window of 3$\times$3 pixels, and a threshold for the magnetic field equal to 10\% of the maximum value it attains above its mean value. We made this choice since the mean value is non-zero in the specific experiment because of the ambient field. Finally, the width of the Gaussian that was convolved with the PIL was taken to equal 9 pixels. The resulting Gaussian mask on the photosphere, $W$, was used to define the helicities,
\begin{equation}
H_\mathrm{x,PIL}=\int_{z=0} h_\mathrm{x}\,W\,{\rm d}\Phi,
\label{flhpil}
\end{equation}
where `x' can be any of the characters `r', `j', or `pj'. These are the parts of the respective helicities contained in the PIL.

\begin{figure}[ht]
\centering
\includegraphics[width=0.46\textwidth]{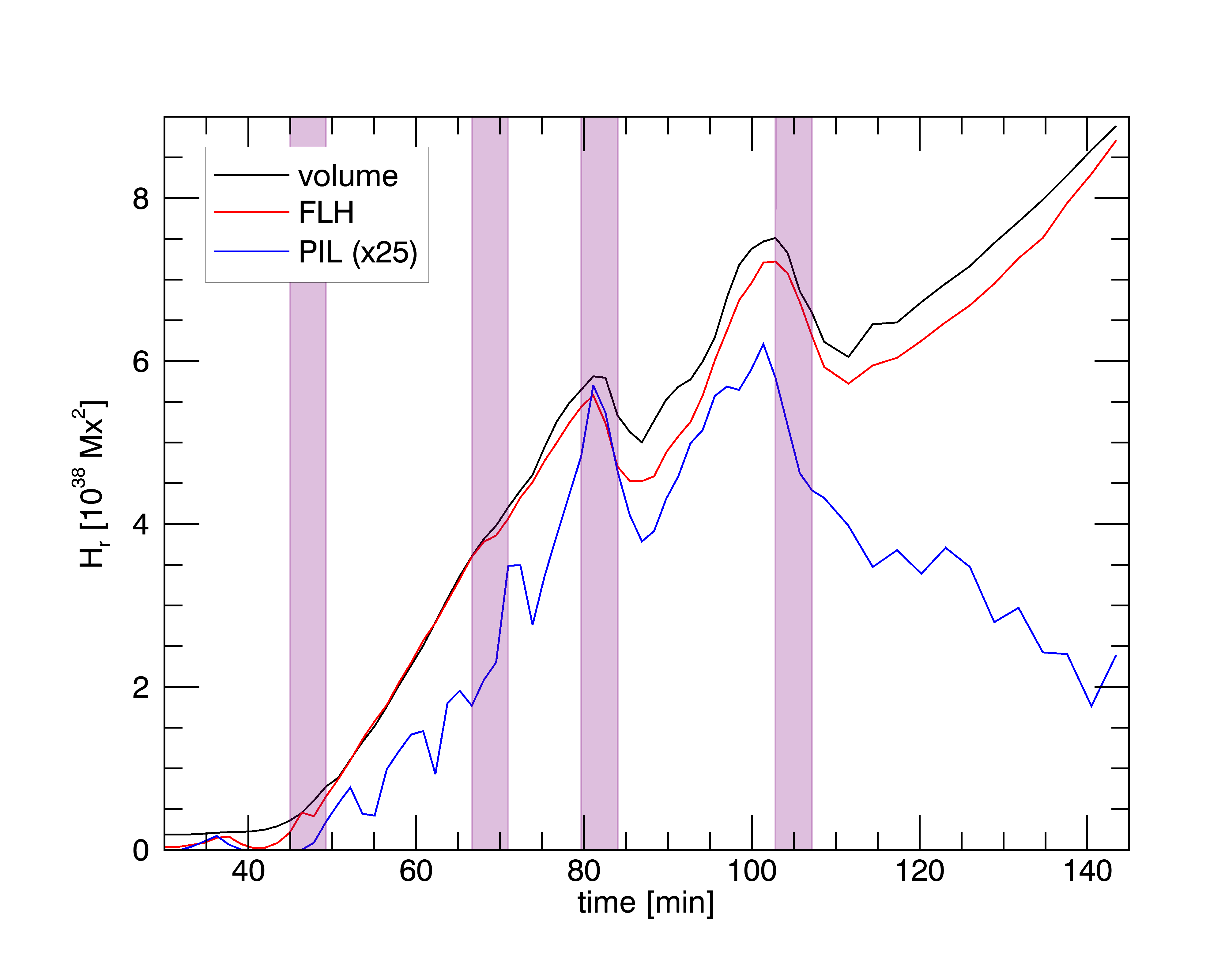}

\includegraphics[width=0.46\textwidth]{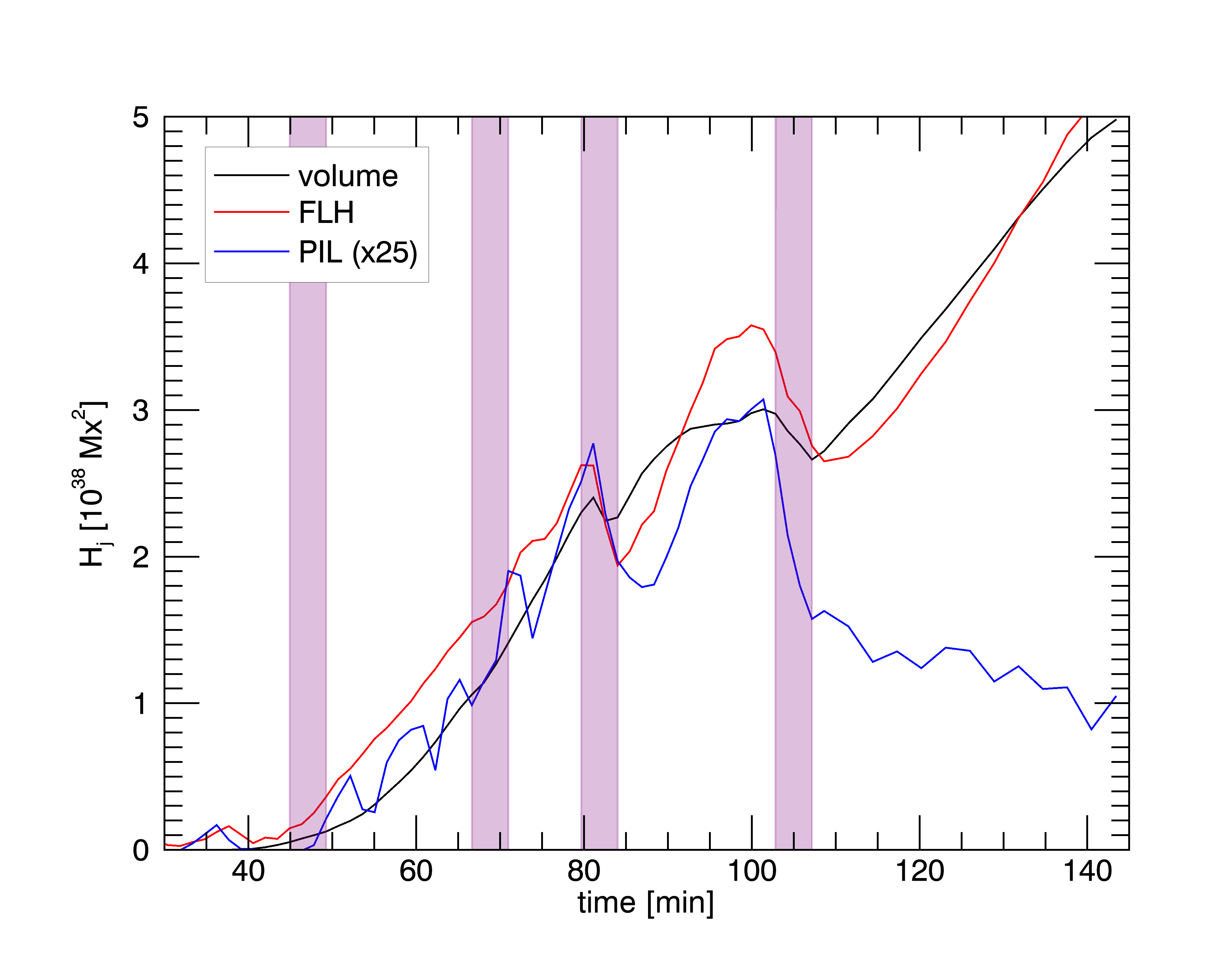}
\caption{(Top) Evolution of the relative helicity computed by the volume method (Eq.~(\ref{helr}), black curve), the helicity computed by the RFLH method (Eq.~(\ref{flhhel}), red curve), and the helicity contained around the PIL (Eq.~(\ref{flhpil}), blue curve).\protect\\
(Bottom) Evolution of the current-carrying helicity computed by the volume method (Eq.~(\ref{helj}), black curve), the respective helicity computed by the RFLH method (Eq.~(\ref{flhhelj}), red curve), and the current-carrying helicity contained around the PIL (Eq.~(\ref{flhpil}), blue curve). The vertical purple stripes represent the time intervals around the four identified jets.}
\label{pilhels}
\end{figure}

In the top panel of Fig.~\ref{pilhels}, we show the evolution of three different relative helicities: $H_\mathrm{r}$ from Eq.~(\ref{helr}), $H_\mathrm{r,fl}$ from Eq.~(\ref{flhhel}), and $H_\mathrm{r,PIL}$ from Eq.~(\ref{flhpil}). We note that the relative helicity derived from the field line helicity (red curve) has a similar evolution pattern as $H_\mathrm{r}$ (black curve). Moreover, their values have absolute relative differences of less than 10\% most of the time. This shows that the RFLH computation method is working as expected and as has also been seen in other cases previously \citep[e.g.][]{moraitis21}. The PIL helicity (blue curve) also follows $H_\mathrm{r}$ until the large blowout jet, despite fluctuating much more and being smaller by a factor of $\sim$25. The fluctuations, which are found in all PIL-related quantities, are caused by the calculation method of the PIL helicities and especially by the number of points comprising the PILs, which vary between snapshots. The jiggling of the PIL relative helicity does not allow us to infer its behaviour during the first two jets, but the peaks of $H_\mathrm{r,PIL}$ near the last two jets coincide with those of $H_\mathrm{r}$. In fact, they are even more pronounced compared to those of the other relative helicities. After the large blowout jet, $H_\mathrm{r,PIL}$ decreases, while the other two helicities increase. This different behaviour of $H_\mathrm{r,PIL}$ indicates that the increase in $H_\mathrm{r}$ after the last jet is due to the coronal field.

In the bottom panel of Fig.~\ref{pilhels}, we show the evolution of the three different current-carrying helicities, $H_\mathrm{j}$, from Eq.~(\ref{helj}), $H_\mathrm{j,fl}$ from Eq.~(\ref{flhhelj}), and $H_\mathrm{j,PIL}$ from Eq.~(\ref{flhpil}). We note that the current-carrying helicity derived from the respective field line helicity (red curve) has similar values to $H_\mathrm{j}$ (black curve), exhibiting relative absolute differences up to 15\%. This slightly higher value compared to the case of the relative helicities (top panel of Fig.~\ref{pilhels}) is mostly due to the difference in the evolution patterns of $H_\mathrm{j,fl}$ and $H_\mathrm{j}$ around $t\sim 100\, \mathrm{min}$, right before the last jet. Similarly to the PIL relative helicity, the PIL current-carrying helicity, $H_\mathrm{j,PIL}$ (blue curve), is also jiggling and identifies the two major peaks during the last two jets more clearly than the respective volume helicity, $H_\mathrm{j}$. Likewise, $H_\mathrm{j,PIL}$ decreases after the large blowout jet, contrary to the other current-carrying helicities. We note that overall the two panels of Fig.~\ref{pilhels} exhibit many similarities between the relations of $H_\mathrm{j}$ and $H_\mathrm{r}$ with their respective PIL helicities.

\section{Discussion}
\label{sect:discussion}

This work examines various energy- and helicity-related quantities in a flux emergence MHD simulation. The presence of the oblique ambient magnetic field in the simulation leads to the production of a number of jets in the coronal volume and to an overall high level of activity. We have identified four jet events as the main focus of our study.

The evolution of the energies and the helicities show jet-related changes in all the identified jets, although these are more pronounced in the last two jets, which are the strongest. The relative helicity and its two components exhibit sharper changes during these jets compared to the energies, while the most jet-indicating among the three energies is the free energy. These remarks strengthen the idea that helicity is a better marker for eruptivity, followed by free energy \citep[e.g.][]{gupta21,duan23}. We stress, however, that both energy and helicity are needed in order to understand the dynamics of a magnetised system. Additionally, an increased transformation of $H_\mathrm{j}$ to $H_\mathrm{pj}$ was observed during the jets, owing to the slightly earlier peaking of $H_\mathrm{j}$, in accordance with the results of \citet{linan18}. 

A noteworthy result of this work is that the eruptivity index shows atypical behaviour in the simulation. Apart from the reconnection jet, when the eruptivity index has a local peak and decreases afterwards temporarily, it shows mixed behaviour during all other jets. A possible reason could be the presence of the ambient field and the resulting increased coronal activity of the specific simulation. It is known, however, that the eruptivity index does not conform to its standard picture in all the observed jets \citep{green22}. Moreover, the eruptivity index shows an intense, broad peak outside the intervals of jet production. It seems that something else happened at that time that did not manifest itself in the temperature images of Fig.~\ref{mhdtemp}, or in any other quantity that we checked.

Another factor that could play an important role in the behaviour of the eruptivity index is the choice of volume in which it is computed, as \citet{lumme22} have shown. The consideration of the individual subvolumes in Sect.~\ref{sect:results3} allowed us to determine how the various profiles change with height. For the eruptivity index, however, this does not show any improvement compared to the case of the whole volume. In contrast, we are able to draw various conclusions for the energies and helicities. The removal of the lower 1~Mm above the photosphere in the first subvolume leads to smoother profiles. The middle volume shows the best agreement with jet activity, in both the energies and the helicities. The energy and helicity variation with height shows that the magnetic field gets more potential and less helical as it goes higher. Our analysis shows that it is important to check the results in different volumes, as extra information can be obtained.

An important aspect that we did not examine in this work was the effect of partial ionisation on our results. We know from \citet{chouliaras23}, for example, that partial ionisation affects the emergence of magnetic flux at the photosphere in a number of ways, and so it could have an impact on our results to some extent. We leave this subject for a future work in which a comparison with the case in which the plasma is fully ionised is going to be made as well.

The main interest of this work is to see the behaviour of the recently introduced PIL relative helicity \citep{moraitis24} in an MHD simulation. The PIL helicity reaffirms the results of that work in a totally different set-up. It seems, therefore, that the relative helicity contained around the PIL is an important factor for studies of solar eruptivity. The PIL helicity is added to the list of other PIL-derived quantities, such as the $R$-parameter of \citet{schrijver07} or the mean twist around the PIL \citep{sun15,li22}, that have been shown to relate to solar activity.

Furthermore, the study, for the first time, of the current-carrying helicity contained around the PIL reveals that it not only agrees qualitatively with the respective volume quantity, but it is even more responsive to jet activity, similarly to the PIL relative helicity. The current-carrying PIL helicity could therefore be equally good at determining upcoming solar eruptivity and so it is worth examining further in the future.

\begin{acknowledgements}
The authors thank the referee for carefully reading the paper and providing constructive comments. This research has received funding from the ERC Whole Sun Synergy grant N$^o$ 810218. GC acknowledges support from the Royal Society grant RGF/EA/180232. The work was supported by the High Performance Computing facilities of the University of St Andrews `Kennedy'.
\end{acknowledgements}

\bibliographystyle{aa}
\bibliography{refs}

\begin{thebibliography}{42}
\expandafter\ifx\csname natexlab\endcsname\relax\def\natexlab#1{#1}\fi

\bibitem[{{Arber} {et~al.}(2001){Arber}, {Longbottom}, {Gerrard}, \&
  {Milne}}]{arber01}
{Arber}, T.~D., {Longbottom}, A.~W., {Gerrard}, C.~L., \& {Milne}, A.~M. 2001,
  J. Comput. Phys., 171, 151

\bibitem[{{Archontis} \& {Hood}(2013)}]{archontis13}
{Archontis}, V. \& {Hood}, A.~W. 2013, \apjl, 769, L21

\bibitem[{{Archontis} {et~al.}(2014){Archontis}, {Hood}, \&
  {Tsinganos}}]{archontis14}
{Archontis}, V., {Hood}, A.~W., \& {Tsinganos}, K. 2014, \apjl, 786, L21

\bibitem[{{Berger}(1988)}]{berger88}
{Berger}, M.~A. 1988, \aap, 201, 355

\bibitem[{{Berger}(1999)}]{berger99}
{Berger}, M.~A. 1999, Plasma Physics and Controlled Fusion, 41, B167

\bibitem[{{Berger} \& {Field}(1984)}]{BergerF84}
{Berger}, M.~A. \& {Field}, G.~B. 1984, J. Fluid. Mech., 147, 133

\bibitem[{{Chouliaras} {et~al.}(2023){Chouliaras}, {Syntelis}, \&
  {Archontis}}]{chouliaras23}
{Chouliaras}, G., {Syntelis}, P., \& {Archontis}, V. 2023, \apj, 952, 21

\bibitem[{{DeVore}(2000)}]{devore00}
{DeVore}, C.~R. 2000, \apj, 539, 944

\bibitem[{Duan {et~al.}(2023)Duan, Jiang, \& Feng}]{duan23}
Duan, A., Jiang, C., \& Feng, X. 2023, \apj, 945, 102

\bibitem[{{Finn} \& {Antonsen}(1985)}]{fa85}
{Finn}, J. \& {Antonsen}, T. 1985, Comments Plasma Phys. Control. Fusion, 9,
  111

\bibitem[{{Green} {et~al.}(2022){Green}, {Thalmann}, {Valori}, {Pariat},
  {Linan}, \& {Moraitis}}]{green22}
{Green}, L.~M., {Thalmann}, J.~K., {Valori}, G., {et~al.} 2022, \apj, 937, 59

\bibitem[{{Gupta} {et~al.}(2021){Gupta}, {Thalmann}, \& {Veronig}}]{gupta21}
{Gupta}, M., {Thalmann}, J.~K., \& {Veronig}, A.~M. 2021, \aap, 653, A69

\bibitem[{{Karpen} {et~al.}(2017){Karpen}, {DeVore}, {Antiochos}, \&
  {Pariat}}]{karpen17}
{Karpen}, J.~T., {DeVore}, C.~R., {Antiochos}, S.~K., \& {Pariat}, E. 2017,
  \apj, 834, 62

\bibitem[{{Leake} \& {Linton}(2013)}]{leake13}
{Leake}, J.~E. \& {Linton}, M.~G. 2013, \apj, 764, 54

\bibitem[{{Li} {et~al.}(2022){Li}, {Sun}, {Hou}, {Chen}, {Yang}, \&
  {Zhang}}]{li22}
{Li}, T., {Sun}, X., {Hou}, Y., {et~al.} 2022, \apjl, 926, L14

\bibitem[{{Linan} {et~al.}(2018){Linan}, {Pariat}, {Moraitis}, {Valori}, \&
  {Leake}}]{linan18}
{Linan}, L., {Pariat}, {\'E}., {Moraitis}, K., {Valori}, G., \& {Leake}, J.
  2018, \apj, 865, 52

\bibitem[{{Lumme} {et~al.}(2022){Lumme}, {Pomoell}, {Price}, {Kilpua},
  {Kazachenko}, {Fisher}, \& {Welsch}}]{lumme22}
{Lumme}, E., {Pomoell}, J., {Price}, D.~J., {et~al.} 2022, \aap, 658, A200

\bibitem[{{Moore} {et~al.}(2010){Moore}, {Cirtain}, {Sterling}, \&
  {Falconer}}]{moore10}
{Moore}, R.~L., {Cirtain}, J.~W., {Sterling}, A.~C., \& {Falconer}, D.~A. 2010,
  \apj, 720, 757

\bibitem[{{Moraitis} {et~al.}(2018){Moraitis}, {Pariat}, {Savcheva}, \&
  {Valori}}]{moraitis18}
{Moraitis}, K., {Pariat}, {\'E}., {Savcheva}, A., \& {Valori}, G. 2018,
  \solphys, 293, 92

\bibitem[{{Moraitis} {et~al.}(2019{\natexlab{a}}){Moraitis}, {Pariat},
  {Valori}, \& {Dalmasse}}]{moraitis19}
{Moraitis}, K., {Pariat}, E., {Valori}, G., \& {Dalmasse}, K.
  2019{\natexlab{a}}, \aap, 624, A51

\bibitem[{{Moraitis} {et~al.}(2021){Moraitis}, {Patsourakos}, \&
  {Nindos}}]{moraitis21}
{Moraitis}, K., {Patsourakos}, S., \& {Nindos}, A. 2021, \aap, 649, A107

\bibitem[{{Moraitis} {et~al.}(2024){Moraitis}, {Patsourakos}, {Nindos},
  {Thalmann}, \& {Pariat}}]{moraitis24}
{Moraitis}, K., {Patsourakos}, S., {Nindos}, A., {Thalmann}, J.~K., \&
  {Pariat}, {\'E}. 2024, \aap, 683, A87

\bibitem[{{Moraitis} {et~al.}(2019{\natexlab{b}}){Moraitis}, {Sun}, {Pariat},
  \& {Linan}}]{moraitis19b}
{Moraitis}, K., {Sun}, X., {Pariat}, {\'E}., \& {Linan}, L. 2019{\natexlab{b}},
  \aap, 628, A50

\bibitem[{{Moraitis} {et~al.}(2014){Moraitis}, {Tziotziou}, {Georgoulis}, \&
  {Archontis}}]{moraitis14}
{Moraitis}, K., {Tziotziou}, K., {Georgoulis}, M.~K., \& {Archontis}, V. 2014,
  \solphys, 289, 4453

\bibitem[{{Moreno-Insertis} \& {Galsgaard}(2013)}]{moreno13}
{Moreno-Insertis}, F. \& {Galsgaard}, K. 2013, \apj, 771, 20

\bibitem[{{Pariat} {et~al.}(2009){Pariat}, {Antiochos}, \& {DeVore}}]{pariat09}
{Pariat}, E., {Antiochos}, S.~K., \& {DeVore}, C.~R. 2009, \apj, 691, 61

\bibitem[{{Pariat} {et~al.}(2017){Pariat}, {Leake}, {Valori}, {Linton},
  {Zuccarello}, \& {Dalmasse}}]{pariat17}
{Pariat}, E., {Leake}, J.~E., {Valori}, G., {et~al.} 2017, \aap, 601, A125

\bibitem[{{Pariat} {et~al.}(2023){Pariat}, {Wyper}, \& {Linan}}]{pariat23}
{Pariat}, E., {Wyper}, P.~F., \& {Linan}, L. 2023, \aap, 669, A33

\bibitem[{{Patsourakos} {et~al.}(2008){Patsourakos}, {Pariat}, {Vourlidas},
  {Antiochos}, \& {Wuelser}}]{patsourakos08}
{Patsourakos}, S., {Pariat}, E., {Vourlidas}, A., {Antiochos}, S.~K., \&
  {Wuelser}, J.~P. 2008, \apjl, 680, L73

\bibitem[{{Raouafi} {et~al.}(2016){Raouafi}, {Patsourakos}, {Pariat}, {Young},
  {Sterling}, {Savcheva}, {Shimojo}, {Moreno-Insertis}, {DeVore}, {Archontis},
  {T{\"o}r{\"o}k}, {Mason}, {Curdt}, {Meyer}, {Dalmasse}, \&
  {Matsui}}]{jetreview}
{Raouafi}, N.~E., {Patsourakos}, S., {Pariat}, E., {et~al.} 2016, \ssr, 201, 1

\bibitem[{{Schrijver}(2007)}]{schrijver07}
{Schrijver}, C.~J. 2007, \apjl, 655, L117

\bibitem[{{Shibata} {et~al.}(1992){Shibata}, {Ishido}, {Acton}, {Strong},
  {Hirayama}, {Uchida}, {McAllister}, {Matsumoto}, {Tsuneta}, {Shimizu},
  {Hara}, {Sakurai}, {Ichimoto}, {Nishino}, \& {Ogawara}}]{shibata92}
{Shibata}, K., {Ishido}, Y., {Acton}, L.~W., {et~al.} 1992, \pasj, 44, L173

\bibitem[{{Sun} {et~al.}(2015){Sun}, {Bobra}, {Hoeksema}, {Liu}, {Li}, {Shen},
  {Couvidat}, {Norton}, \& {Fisher}}]{sun15}
{Sun}, X., {Bobra}, M.~G., {Hoeksema}, J.~T., {et~al.} 2015, \apjl, 804, L28

\bibitem[{{Syntelis} {et~al.}(2017){Syntelis}, {Archontis}, \&
  {Tsinganos}}]{syntelis17}
{Syntelis}, P., {Archontis}, V., \& {Tsinganos}, K. 2017, \apj, 850, 95

\bibitem[{{Thalmann} {et~al.}(2019{\natexlab{a}}){Thalmann}, {Linan}, {Pariat},
  \& {Valori}}]{thalmann19a}
{Thalmann}, J.~K., {Linan}, L., {Pariat}, E., \& {Valori}, G.
  2019{\natexlab{a}}, \apjl, 880, L6

\bibitem[{{Thalmann} {et~al.}(2019{\natexlab{b}}){Thalmann}, {Moraitis},
  {Linan}, {Pariat}, {Valori}, \& {Dalmasse}}]{thalmann19}
{Thalmann}, J.~K., {Moraitis}, K., {Linan}, L., {et~al.} 2019{\natexlab{b}},
  \apj, 887, 64

\bibitem[{{T{\"o}r{\"o}k} {et~al.}(2024){T{\"o}r{\"o}k}, {Linton}, {Leake},
  {Miki{\'c}}, {Lionello}, {Titov}, \& {Downs}}]{torok24}
{T{\"o}r{\"o}k}, T., {Linton}, M.~G., {Leake}, J.~E., {et~al.} 2024, \apj, 962,
  149

\bibitem[{{Valori} {et~al.}(2012){Valori}, {D{\'e}moulin}, \& {Pariat}}]{val12}
{Valori}, G., {D{\'e}moulin}, P., \& {Pariat}, {\'E}. 2012, \solphys, 278, 347

\bibitem[{{Valori} {et~al.}(2020){Valori}, {D{\'e}moulin}, {Pariat}, {Yeates},
  {Moraitis}, \& {Linan}}]{valori20}
{Valori}, G., {D{\'e}moulin}, P., {Pariat}, E., {et~al.} 2020, \aap, 643, A26

\bibitem[{{Valori} {et~al.}(2016){Valori}, {Pariat}, {Anfinogentov}, {Chen},
  {Georgoulis}, {Guo}, {Liu}, {Moraitis}, {Thalmann}, \& {Yang}}]{valori16}
{Valori}, G., {Pariat}, {\'E}., {Anfinogentov}, S., {et~al.} 2016, \ssr, 201,
  147

\bibitem[{{Woltjer}(1958)}]{woltjer58}
{Woltjer}, L. 1958, Proceedings of the National Academy of Science, 44, 489

\bibitem[{{Yeates} \& {Page}(2018)}]{yeates18}
{Yeates}, A.~R. \& {Page}, M.~H. 2018, Journal of Plasma Physics, 84, 775840602

\end{thebibliography}

\end{document}